\documentclass[%
 aip,
 amsmath,amssymb,
 reprint,%
]{revtex4-1}

\usepackage{graphicx}
\usepackage{dcolumn}
\usepackage{bm}

\usepackage[utf8]{inputenc}
\usepackage[T1]{fontenc}
\usepackage{mathptmx}
\usepackage{etoolbox}

\makeatletter
\def\@email#1#2{%
 \endgroup
 \patchcmd{\titleblock@produce}
  {\frontmatter@RRAPformat}
  {\frontmatter@RRAPformat{\produce@RRAP{*#1\href{mailto:#2}{#2}}}\frontmatter@RRAPformat}
  {}{}
}%
\makeatother
\begin{document}

\preprint{AIP/123-QED}

\title{Synthesis and Anisotropic Magnetism in Quantum Spin Liquid Candidates $A$YbSe$_2$ ($A$ = K and Rb)}
\author{Jie Xing}
\affiliation{%
Materials Science and Technology Division, Oak Ridge National Laboratory, Oak Ridge, Tennessee 37831, USA
}%
\author{Liurukara D. Sanjeewa}
\affiliation{%
Materials Science and Technology Division, Oak Ridge National Laboratory, Oak Ridge, Tennessee 37831, USA
}%
\affiliation{Missouri Research Reactor, University of Missouri, Columbia, Missouri 65211, USA}
\affiliation{Department of Chemistry, University of Missouri, Columbia, MO 65211, USA}
\author{Andrew F. May}
\affiliation{%
Materials Science and Technology Division, Oak Ridge National Laboratory, Oak Ridge, Tennessee 37831, USA
}%
\author{Athena S. Sefat}
\affiliation{%
Materials Science and Technology Division, Oak Ridge National Laboratory, Oak Ridge, Tennessee 37831, USA
}%
\date{\today}

\begin{abstract}
Quantum spin liquid (QSL) state in rare-earth triangular lattice has attracted much attention recently due to its potential application in quantum computing and communication. Here we report the single-crystal growth synthesis, crystal structure characterizations and magnetic properties of $A$YbSe$_{2}$ ($A$= K, Rb) compounds. The X-ray diffraction analysis shows that $A$YbSe$_{2}$ ($A$ = K and Rb) crystallizes in a trigonal space group, $R$-3$m$ (No. 166) with Z = 3. $A$YbSe$_{2}$ possesses a two-dimensional (2D) Yb-Se-Yb layered structure formed by edged-shared YbSe$_6$ octahedra. The magnetic properties are highly anisotropic for both title compounds and no long-range order is found down to 0.4 K, revealing the possible QSL ground state in these compounds. The isothermal magnetization exhibits one-third magnetization plateau when the magnetic fields are applied in $ab$-plane. Heat capacity is performed along both $ab$-plane and $c$-axis, and features the characteristic dome for triangular magnetic lattice compounds as a function of magnetic fields. Due to the change of the interlayer and intralayer distance of Yb$^{3+}$, the dome shifts to low fields from KYbSe$_2$ to RbYbSe$_2$. All these results indicate the $A$YbSe$_2$ family presents unique frustrated magnetism close to the possible QSL and noncolinear spin states.
\end{abstract}

\maketitle

\section{\label{sec:level1}Introduction}

The quantum spin liquid (QSL) state in geometric frustrated materials attracts wide interests due to the future application of quantum entanglement.~\cite{sadoc_mosseri_1999,Moessner} The highly entangled spins in frustrated magnetic lattices prevent the breaking of any symmetry, even at zero temperature.~\cite{Anderson,Mila2000,Balents} Besides the proposed quantum magnets with spin-1/2 transition metal, the rare-earth based materials are also a rich ground for frustrated magnetism.~\cite{RevModPhys,Broholmeaay0668,li2020spin,sarkis2020unravelling,pajerowski2020quantification,rau2018frustration,xingYbCl3} As an example, Yb-based triangular magnetic lattice structures have been proposed as effective spin $S_{\mathrm{eff}}$ = 1/2 materials at low temperature due to the crystal electric field (CEF) effect.~\cite{Wu1,Wu2,agrapidis,Hester,lispinon,lieffect,lianisotropic} One Yb$^{3+}$ based triangular lattice material, YbMgGaO$_4$, has been proposed as a QSL candidate by the magnetization, heat capacity, neutron scattering, muon spin relaxation and thermal conductivity measurements.~\cite{li2015gapless,li2015rare,li2016muon,shen2016evidence,paddison2017continuous,zhang2018hierarchy,li2017crystalline,steinhardt2019field,li2019ybmggao4} YbMgGaO$_4$ crystallizes in high symmetry (trigonal $R$-3$m$) by creating a perfect Yb$^{3+}$ triangular magnetic lattice. Non-magnetic Mg$^{2+}$ and Ga$^{3+}$ ions separate the Yb$^{3+}$ triangular magnetic layers by 8.4 \AA$~$and the distance between the two Yb$^{3+}$ ions within the layer is 3.4 \AA, which provides well 2D character for YbMgGaO$_4$. However, due to the similar ionic radii of Mg$^{2+}$ and Ga$^{3+}$, both Mg$^{2+}$ and Ga$^{3+}$ tend to mix occupy in the structure. Since Mg$^{2+}$ and Ga$^{3+}$ have two different charges, this can affect the crystal electric field splitting of the Yb$^{3+}$ site compared to the perfect structure. Consequently, this type of intrinsic Mg/Ga disorder between the Yb$^{3+}$ layers could perturb the spin liquid ground state.~\cite{li2015gapless,paddison2017continuous,zhang2018hierarchy,shen2018fractionalized,zhu2017disorder,zhu2018topography,kimchi2018valence} Therefore, it is important not only to study new Yb$^{3+}$ based 2D geometrically frustrated materials, but also to develop rational synthetic pathways targeting novel Yb-based 2D compounds without site disorder, particularly ones that enable the effect of QSL ground state by varying the intralayer and interlayer distances of the Yb$^{3+}$ magnetic lattice.

Among the most recently studied Yb$^{3+}$ triangular magnetic lattices, the delafossite structure (112-structure type) $ARQ_2$ ($A$ = Li, Na, K, Rb, Cs, Tl, Ag, Cu; $R$ = rare earth; $Q$ = O, S, Se, Te) attracts widespread attention in search of QSL candidates.~\cite{liu2018rare,xing2019,jie2019naerse,zangeneh2019single,scheie2020crystal,bordelon2021frustrated} The high symmetry of $ARQ_2$ structures is reported to have different packing structures, where rare earth ions form 2D triangular lattice. However, depending upon the $A$-site cation, $ARQ_2$ structure could crystallize in different space groups.~\cite{xing2019} Further, magnetic properties of $ARQ_2$ compounds can be systematically tuned by either changing $A$-site cation (interlayer-interactions), number of 4$f$ electrons in the rare earth ions or by changing the type of $Q$-site (intralayer interactions). Therefore, $ARQ_2$ structures serve as an experimental touchstone to identify novel quantum magnetic ground states such as QSL.~\cite{liu2018rare,xing2019,jie2019naerse,zangeneh2019single,scheie2020crystal,gao2020crystal,xing2021stripe} The presence of $S_{\mathrm{eff}}$ = 1/2 ground state of Yb$^{3+}$ based NaYb$Q_2$ structures at low temperature region is proposed by electron spin resonance, magnetization and neutron scattering measurements.\cite{baenitz2018naybs,bordelon2020spin,zhang2021crystalline,sichelschmidt2020effective,guo2020magnetic,zhang2021effective,schmidt2021Yb} Further, no long-range order has been observed in heat capacity at zero field down to mK temperatures at low magnetic fields.~\cite{liu2018rare,ranjith2019field,Ranjith2019naybse,PhysRevB.100.220407,ding2019gapless,ma2020spin,ferreira2020frustrated,liybs2,iizuka2020single,zhang2020crystalline} Quantum disordered or QSL ground state is proposed in NaYbO$_2$~\cite{ranjith2019field,ding2019gapless,bordelon2019field} and the signature of the magnetic excitation continuum has been found in NaYbSe$_2$ from 0.1 to 2.5 meV in inelastic neutron scattering (INS) measurement, suggesting the ground state of NaYbSe$_2$ is a QSL with a spinon Fermi surface.~\cite{dai2021spinon} The continuum-like excitation with short-range order has been found in CsYbSe$_2$, indicating it may be located close to the critical boundary between QSL and 120$^{\circ}$ phase.\cite{xie2021field} The external magnetic fields could perturb the zero-field magnetic ground state and induce the long-range order where the characteristic 1/3 magnetization plateau can be observed in isothermal magnetization.~\cite{Ranjith2019naybse,PhysRevB.100.220407} These field-induced phase transitions reveal the unique magnetic phase diagram in Yb$^{3+}$ based 112 compounds.~\cite{maksimov2019anisotropic,yamamoto2014quantum} Moreover, superconductivity has been reported in NaYbSe$_2$ (T$_c$ $\sim$8 K) under high pressure.~\cite{jia2020mott,zhang2020pressure}

These variations in the structures and the magnetic properties have motivated us to undertake a comprehensive study on $A$YbSe$_2$ ($A$ = K, Rb, Cs), where large $A$ ions could avoid the possible mixed occupancy of $A$/Yb. In the first step of our investigation in this series, we have studied the magnetic properties of CsYbSe$_2$ using ac and dc magnetic susceptibility and heat capacity down to 0.4 K.\cite{xing2019,PhysRevB.100.220407} Additionally, our INS on CsYbSe$_2$ single crystals suggest that this system could provide a natural realization of QSL ground state\cite{xie2021field}. For this purpose, we extend our study to investigate structure-property correlation of KYbSe$_2$ and RbYbSe$_2$.  
In this paper, we discuss the single crystal growth, structure characterization and magnetic properties of KYbSe$_2$ and RbYbSe$_2$. To our knowledge, there have not been any reported studies of the magnetic properties of KYbSe$_2$ and RbYbSe$_2$, however, synthesis and the crystal structure of KYbSe$_2$ has been reported by Gray et al.~\cite{gray2003crystal} Single crystal X-ray confirmed that both KYbSe$_2$ and RbYbSe$_2$ crystallizes in $R$-3$m$ space group. The magnetization and heat capacity are performed down to 0.4 K when the applied magnetic field is along the $ab$-plane and $c$-axis. Both KYbSe$_2$ and RbYbSe$_2$ do not exhibit long-range ordering above 0.4 K suggesting the QSL ground state. Field-induced magnetic transitions are found, and heat capacity form a dome shape in the $H$-$T$ phase diagram. The 1/3 magnetization plateau is observed in both KYbSe$_2$ and RbYbSe$_2$ when magnetic field is applied along the $ab$ plane. Finally, this study allows us to establish the empirical relationship between the intralayer/interlayer space of Yb$^{3+}$ ions and the magnetic properties of $A$YbSe$_2$ ($A$ = Na, K, Rb and Cs) family of compounds.

\section{\label{sec:level1}EXPERIMENTAL SECTION}

\begin{figure}[tbh]
\includegraphics[width=1\linewidth]{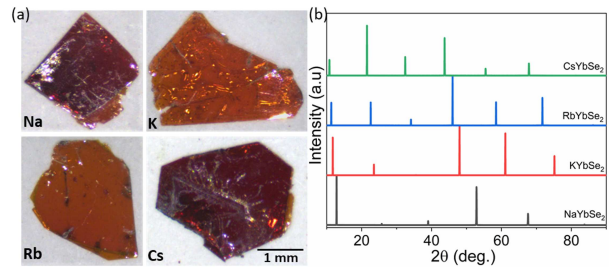}
\caption {\label{fig:crys}(a) Single crystals of $A$YbSe$_2$. (b) (00l) peaks of X-ray diffraction on the flat surface ($ab$-plane) of $A$YbSe$_2$ single crystals.}
\end{figure}

Millimeter-sized hexagonal shape KYbSe$_2$ and RbYbSe$_2$ single crystals, synthesized by a two-step method, are shown in Fig.~\ref{fig:crys}(a). In the first step, K or Rb chunk (99.9\%, Alfa Aesar), and Yb ingot (99.9\%, Alfa Aesar) and Se shot (99.999\%, Alfa Aesar) were loaded by the stoichiometric ratio into an alumina crucible and sealed it in a quartz tube. The ampoules were slowly heated to 850 $^{\circ}$C and cooled down to room temperature in the furnace. After the first reaction, a red color powder was recovered from the crucibles. In the second step, the resulting powder was loaded into quartz tube with mass ratio 1:10 corresponding flux (KCl or RbCl). After that, tubes were sealed again. The final reaction was 850 $^{\circ}$C for 2 weeks. Finally, the reactions were quenched by water. The red plate-like single crystals (Fig.~\ref{fig:crys}(a)) were collected by washing with de-ionized water using suction filtration method. A similar method was employed to synthesize single crystals of both NaYbSe$_2$ and KLuSe$_2$, which were used as references in our measurements. 

Energy-dispersive spectroscopy analysis (EDS) was performed using a Hitachi S-3400 scanning electron microscope equipped with an OXFORD EDX microprobe to identify the elemental composition in single crystals and it confirmed the molar ratio of K(Rb):Yb:Se is 1:1:2 as expected. 
X-ray single-crystal diffraction (SXRD) data were collected using Bruker Quest D8 single-crystal X-ray diffractometer equipped with Mo K$_\alpha$ radiation, $\lambda$ = 0.71073 \AA. The data were collected at 200 K using a nitrogen cold stream and crystals were mounted in paratone oil. The crystal diffraction images were collected using $\phi$ and $\omega$-scans. The diffractometer was equipped with an Incoatec I$\mu$S source using the APEXIII software suite for data setup, collection, and processing. ShelX was utilized for the structural solution and refinement after data reduction via SMART-Plus; spherical absorptions corrections were performed within ShelX after initial refinement.  Twinning was taken into account during refinement.~\cite{sheldrick2015crystal} Table~\ref{tab:crys} provides the detailed unit cell parameters of $A$YbSe$_2$ series and the selected bond distances are summarized in Table~\ref{tab:angle}. X-ray powder diffraction (PXRD) data was collected on a PANalytical X'pert Pro diffractometer equipped with an incident beam monochromator (Cu K$_{\alpha1}$ radiation) at room temperature. Diffraction patterns of the flat surface of $A$YbSe$_2$ crystals were collected to confirm the 00$l$ orientation ($ab$-plane), as shown in Fig.~\ref{fig:crys}(b).

Magnetic properties were measured using a Quantum Design (QD) Magnetic Properties Measurement System (MPMS3). The magnetization below 2 K was measured by the MPMS3 iHe3 option. The total mass of the measured crystals was $\approx$3 mg. Temperature dependent heat capacity was measured in QD Physical Properties Measurement System (PPMS) using the relaxation technique. The mass of the single crystals for heat capacity was $\approx$1 mg.

\begin{table*}[tbh]
\caption{\label{tab:crys}Crystallographic data of $A$YbSe$_2$ determined by single-crystal X-ray diffraction. }

\begin{tabular}{|p{4cm}|p{3cm}|p{3cm}|p{3cm}|p{3cm}|}
\hline
empirical formula&	NaYbSe$_2$~\cite{liu2018rare}	&KYbSe$_2$	&RbYbSe$_2$	&CsYbSe$_2$~\cite{xing2019}\\
\hline
formula weight (g/mol)&	441.18&	370.06&	416.43&	463.87\\
\hline
$T$, K&	298&	200	&200&	298\\
\hline
crystal habitat&	red plates&	red plates	&red plates	&red plates\\
\hline
crystal dimensions, mm&	0.08$\times$0.02$\times$0.02&	0.04$\times$0.02$\times$0.02&	0.04$\times$0.02$\times$0.02	&0.04$\times$0.02$\times$0.02\\
\hline
crystal system&	trigonal&	trigonal&	trigonal&	hexagonal\\
\hline
space group&	$R$-3$m$ (No.166)&	$R$-3$m$ (No.166)&	$R$-3$m$ (No.166)&	$P$6$_3$/$mmc$ (No.194)\\
\hline
$a$, \AA&	4.0568(8)&	4.1149(5)&	4.1371(1)&	4.1484(4)\\
\hline
$c$, \AA&	20.772(6)&	22.6911(4)&	23.6428(5)	&16.549(3)\\
\hline
volume, \AA$^3$&	296.06	&332.76(2)&	350.45(2)&	246.64(7)\\
\hline
$Z$&	3&	3&	3&	2\\
\hline
$D$ (calc), g/cm$^3$&	–&	5.540&	5.920&	6.246\\
\hline
$\mu$ (Mo K$_\alpha$), mm$^{-1}$	&–&	38.200&	45.730&	40.788\\
\hline
$F$(000)&	–&	471	&525&	386\\
\hline
$T_\mathrm{max}$, $T_\mathrm{min}$&	–&	0.1576-1.0000&	0.1986-1.0000&	0.1702-1.0000\\
\hline
$\theta$ range&	–&	5.392-38.789&	2.58-38.72&	2.46-30.40\\
\hline
reflections collected&	–&	3934&	4222&	5806\\
\hline
data/restraints/parameters&	–&	456/0/10&	485/0/10&	116/0/9\\
\hline
final $R$ [$I$> 2$\sigma$($I$)] $R_\mathrm{1}$, $R_\mathrm{w2}$&	–	&0.0246/0.660&	0.0178/0.0402&	0.0549/0.01427\\
\hline
final $R$ (all data) $R_\mathrm{1}$, $R_\mathrm{w2}$&	–&	0.0251/0.0663&	0.0210/0.0424&	0.0552/0.1428\\
\hline
GoF	&–&	1.078&	1.092	&1.023\\
\hline
largest diff. peak/hole, e/\AA$^3$&	–	&1.454/-2.3631&	3.401/-2.911&	1.985/-3.949\\
\hline
\end{tabular}
\label{Tab1}
\end{table*}

\section{\label{sec:level1}RESULTS AND DISCUSSION}
\subsection{\label{sec:level2}Crystal Structure}
\begin{figure}[tbh]
\includegraphics[width=1\linewidth]{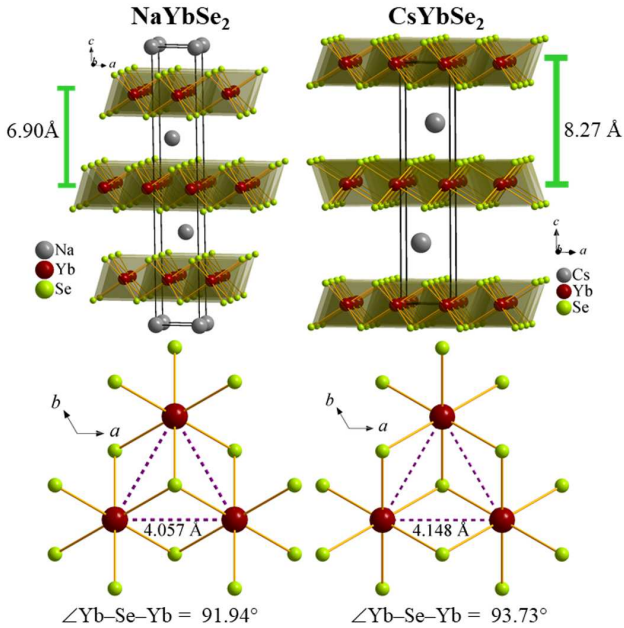}
\caption {\label{fig:struc}Crystal structure comparison between the $R$-3$m$ in NaYbSe$_2$ and $P$6$_3$/$mmc$ in CsYbSe$_2$. In $A$YbSe$_2$ ($A$ = Na, K, Rb, Cs) series, replacing Na with larger cations such as K, Rb, and Cs changes the intralayer distances, Yb–Se–Yb bond angles, and the interlayer distances.}
\label{crystal}
\end{figure}

Single crystal X-ray structure investigation reveals that both KYbSe$_2$ and RbYbSe$_2$ adopt the trigonal $R$-3$m$ space group compare to $P$6$_3$/$mmc$ in CsYbSe$_2$. The detailed crystallographic information of $A$YbSe$_2$ ($A$ = Na, K, Rb, Cs) is given in Table~\ref{tab:crys}. The crystal structure is built from 2D Se–Yb–Se layers where $A$-site cations reside in between the layers. These 2D Se–Yb–Se layers packed along the $c$-axis as displayed in Fig.~\ref{fig:struc}, however, the packing of Se–Yb–Se layers and $A$-site cations are different between the $R$-3$m$ and $P$6$_3$/$mmc$ space groups. The interlayer distance is mainly controlled by the size of the $A$-site cation. Table~\ref{tab:angle} summarizes the bond distances and angles of $A$YbSe$_2$ series of compounds, including the interlayer distances. For example, in NaYbSe$_2$, the interlayer distance is 6.97 \AA $~$while in CsYbSe$_2$ it is 8.24 \AA. This is mainly due to the larger size of the Cs$^+$ ion compared to the Na$^+$ ion. The interlayer distances for KYbSe$_2$ and RbYbSe$_2$ are 7.56 and 7.80 \AA, respectively. The 2D Yb–Se–Yb layers are constructed along the $ab$-plane from edged sharing YbSe$_6$ units, as displayed in Fig.~\ref{fig:struc}. In both structure types, Yb$^{3+}$ ions form an equilateral triangle with Se$^{2-}$ sitting in the threefold symmetry. It is noteworthy to mention that the size of the Yb$^{3+}$ equilateral triangle represent the $a$-axis of the structures and this is the intralayer distance between the Yb$^{3+}$ ions within the 2D lattice. As pointed in Fig.~\ref{fig:struc}, there is a significant variation between the intralayer distances between NaYbSe$_2$ and CsYbSe$_2$. It is 4.057 and 4.148 \AA$~$ for NaYbSe$_2$ and CsYbSe$_2$, respectively. Additionally, Yb–Se–Yb bond angle also changes significantly throughout the $A$YbSe$_2$ series. As pointed in Fig.~\ref{fig:struc}, Yb–Se–Yb bond angles are 91.94$^{\circ}$ and 93.73$^{\circ}$ for NaYbSe$_2$ and CsYbSe$_2$, respectively. These subtle differences in intra- and interlayer distances, and bond angles clearly effect the magnetic properties of this series of compounds, as we discuss in the next section.

\begin{table}[tbh]
\caption{\label{tab:angle}The selected bond distances (\AA) and angles ($^\circ$) of $A$YbSe$_2$ series.}

\begin{tabular}{|p{2.3cm}|p{1.45cm}|p{1.45cm}|p{1.45cm}|p{1.45cm}|}
\hline
& NaYbSe$_2$~\cite{liu2018rare}&	KYbSe$_2$&	RbYbSe$_2$	&CsYbSe$_2$~\cite{xing2019}\\
\hline
Yb‒Se&	4.0568&	4.1149(5)&	4.1371(1)	&4.1484(4)\\
\hline
Yb‒Se‒Yb&	91.14&	93.13(2)&	93.53(2)&	93.77(2)\\
\hline
Interlayer Yb‒Yb&	6.90&	7.56&	7.80&	8.27\\
\hline

\end{tabular}
\label{Tab1}
\end{table}

\begin{table}[tbh]
\caption{\label{tab:fitting}The effective magnetic moment ($\mu_\mathrm{eff}$) and Curie-Weiss temperature ($\theta_\mathrm{CW}$), and Van Vleck term $\chi_\mathrm{vv}$ of KYbSe$_2$ and RbYbSe$_2$ obtained from the fit at high temperatures (HT) of $200-250$ K, and at low temperatures (LT) of $4-10$ K.}
\begin{ruledtabular}
\begin{tabular}{cccccc}
 & $\mu_{\mathrm{eff}}$[$\mu_{\mathrm{B}}$] & $\theta_{CW}$[K] & $\mu_{\mathrm{eff}}$[$\mu_{\mathrm{B}}$] & $\theta_{CW}$[K] &$\chi_\mathrm{vv}$[emu/mol Oe] \\
 & HT & HT & LT & LT&LT \\
\hline
KYbSe$_2$\\
$H\|$\emph{ab}  & 4.87  &  -73.5 & 3.41 &-12.6& 0.002\\
$H\|$\emph{c}   & 5.08  &  -40.3 & 0.65  & -4.7&0.027 \\
\hline
RbYbSe$_2$\\
$H\|$\emph{ab}  & 4.44  &  -57.9 & 3.45 &-15.0& 0.005\\
$H\|$\emph{c}   & 5.01  &  -41.9 & 0.67  & -4.5&0.026 \\
\end{tabular}
\end{ruledtabular}
\end{table}
\subsection{\label{sec:level2}Magnetic Properties of $A$YbSe$_2$}

\begin{figure}[tbh]
\includegraphics[width=1\linewidth]{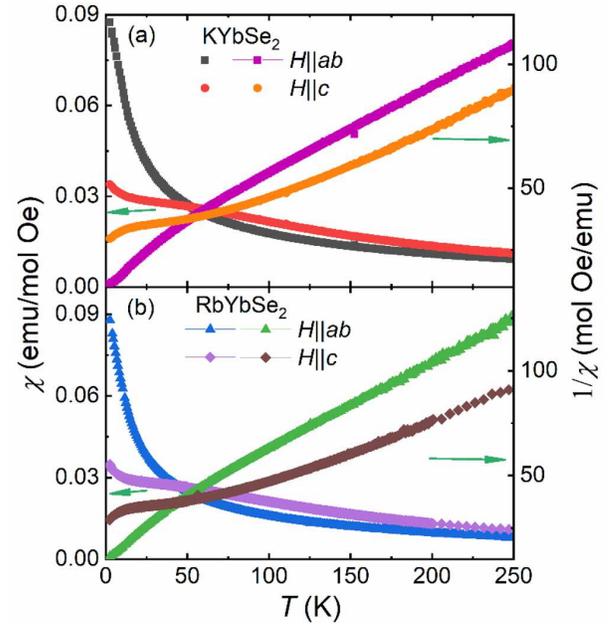}
\caption {\label{fig:magnetization}Temperature dependence of magnetic susceptibility of (a) KYbSe$_2$ or (b) RbYbSe$_2$ below 1.6 K, from 0.1 T to 7 T applied magnetic fields. }
\label{crystal}
\end{figure}

The geometrical frustrated Yb$^{3+}$ triangular lattices with different intralayer and interlayer distances inspire us to investigate the relationship between the Yb-based 112-typed structures and their magnetism. The orientation is determined by the flat surface, where sharp 00l peaks present in Fig.~\ref{fig:crys}(b). Fig.~\ref{fig:magnetization}(a-b) present the temperature dependence of magnetic susceptibility $\chi$ =$M$/$H$, where $M$ is the magnetization and $H$ the applied field, and the inverse magnetic susceptibility is also shown above 2 K along $ab$-plane and $c$-axis under 1 T applied magnetic field. No long-range order or anomalies are observed above 2 K. Large anisotropy between $H$||$ab$ and $H$||$c$ appears below 50 K, with a magnitude ratio of approximately 3 at 2 K. The temperature dependence of 1/$\chi$ exhibits a linear relation at high temperature. When the temperature decreases below 100 K, the curves strongly bend on both directions in KYbSe$_2$ and RbYbSe$_2$, indicating the systems go to the low spin state of Yb$^{3+}$ due to the CEF effect. The Curie-Weiss fit of the magnetization at 200 K to 250 K is used in both directions for the two compounds. The Curie Weiss fit including a temperature independent Van Vleck term is used at low temperature region (4-10 K). We present the fitting parameters in Table~\ref{tab:fitting}. The fitting results show comparable value in KYbSe$_2$ and RbYbSe$_2$ on each direction, indicating the similar magnetic interaction of Yb$^{3+}$ ions in these two structures. The negative Curie-Weiss temperatures in all measurements indicate the antiferromagnetic interactions between Yb$^{3+}$. The effective moment from high temperature fitting is close to the moment of free Yb$^{3+}$ (4.54 $\mu_{\mathrm{B}}$). At low temperature, the effective moments present anisotropic smaller values along these two directions when we consider Van Vleck contribution. Although the precious g values need further electron spin resonance and high magnetic field measurements to confirm, these fitting results at low temperature regions hint at the possible anisotropic g values in these two compounds.

\begin{figure}[tbh]
\includegraphics[width=1\linewidth]{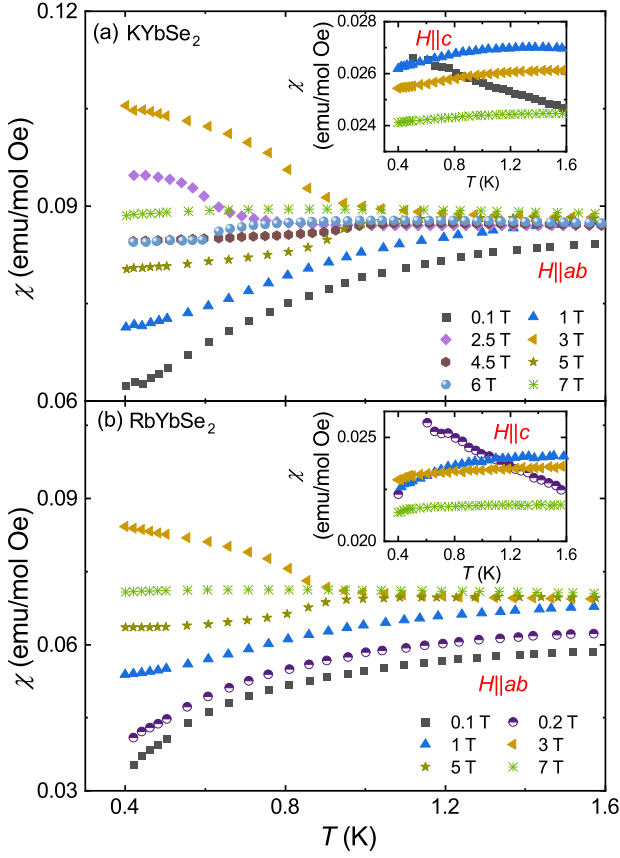}
\caption {\label{fig:he3magne}Temperature dependence of magnetic susceptibility and inverse magnetic susceptibility for (a) KYbSe$_2$ and (b) RbYbSe$_2$.}
\label{crystal}
\end{figure}

The low-T magnetization data of KYbSe$_2$ is shown in Fig.~\ref{fig:he3magne}(a) and no evidence of long-range order is observed. However, a broad maximum is found below 2 K, suggesting short-range interactions. This feature is similar to previous report for CsYbSe$_2$ and NaYbSe$_2$, but different from NaYbS$_2$.~\cite{ranjith2019field,Ranjith2019naybse,PhysRevB.100.220407} The maximum becomes broader with increasing the magnetic field. A small jump appears at $\sim$0.7 K at 2.5 T, which reveals the field-induced long-range magnetic order. The transition temperature increases to maximum $\sim$1 K with field up to 3 T, then decreases to 0.7 K at 6 T, and finally disappears above 0.4 K at 7 T. The inset of Fig.~\ref{fig:he3magne}(a) presents the temperature dependence of magnetization under different fields along $c$-axis. At the low field, the magnetization increases monotonously by decreasing the temperature. However, a maximum appears from 1 to 7 T, suggesting short-range interactions. Due to the large interlayer distance, the long-range order needs a higher magnetic field than our instrument limit of 7 T. The magnitude at $H$||$c$ is much smaller than that at $H$||$ab$, which proves XY-like interactions between Yb$^{3+}$. These low temperature magnetization results indicate that KYbSe$_2$ has the possible spin liquid ground state with the effective spin 1/2 state at zero field, and magnetic fields could induce magnetic transitions above 2 T at $H$||$ab$. A similar result of RbYbSe$_2$ is presented in Fig.~\ref{fig:he3magne}(b), besides the detailed transition temperature shifts to low temperature due to the slight expansion in the structure. 

\begin{figure}[tbh]
\includegraphics[width=1\linewidth]{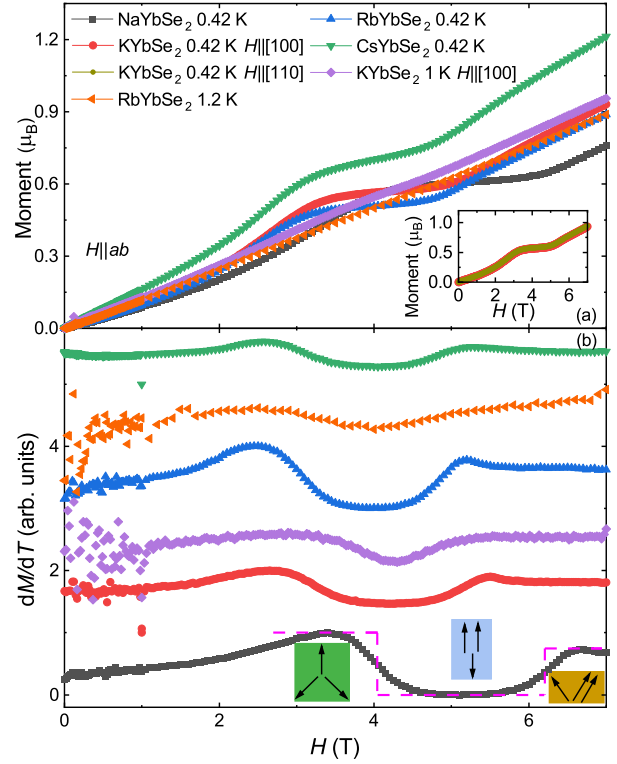}
\caption {\label{fig:isomagnetization}(a) Isothermal magnetization with $H$||$ab$ for $A$YbSe$_2$ at 0.42 K, KYbSe$_2$ at 1 K, and RbYbSe$_2$ at 1.2 K. The inset shows the overlapping isothermal magnetization of KYbSe$_2$ at $H$||[100] and $H$||[110], at 0.42 K. (b) The differential of magnetization versus field of $A$YbSe$_2$. The curves are offset to present clearly.}
\label{crystal}
\end{figure}

The isothermal magnetization of $A$YbSe$_2$ ($A$= Na, K, Rb, Cs) is shown in Fig.~\ref{fig:isomagnetization}(a). We could observe all four samples present the similar 1/3 magnetic plateau when magnetic fields are applied in the easy-plane ($ab$-plane). From the theoretical calculations, the 1/3 magnetic plateau is expected in the easy-plane XXZ model in the quantum picture.~\cite{yamamoto2014quantum} The inset of Fig.~\ref{fig:isomagnetization}(a) presents the isothermal magnetization of KYbSe$_2$ at $H$||[100] and $H$||[110] at 0.42 K. No clear difference is found in these two directions. To show the phase boundaries more clearly, we plot the d$M$/d$H$ vs $H$ in Fig.~\ref{fig:isomagnetization}(b). The pink dash line shows the boundary between three states near the plateau. The plateau shifts to the lower fields from NaYbSe$_2$ to CsYbSe$_2$ due to the change of intralayer and interlayer interaction. NaYbSe$_2$ requires the highest magnetic field to excite the plateau at 0.42 K and among the four samples, as it has the shortest distances between Yb$^{3+}$ in the triangular layers. KYbSe$_2$ and RbYbSe$_2$ follow with the same relation, however, the difference between CsYbSe$_2$ and RbYbSe$_2$ is very small. This could be due to the small differences between the interlayer distances of the Rb and CsYbSe$_2$ structures. Also, we cannot ignore the change of the Yb–Se–Yb stacking sequence between the two structures, which could affect the local environment of YbSe$_6$ octahedra. Moreover, the two-peak feature in d$M$/d$H$ of KYbSe$_2$ and RbYbSe$_2$ can be observed even at $\sim$1 K, suggesting the strong spin fluctuation above the long-range order temperature.

\begin{figure}[tbh]
\includegraphics[width=1\linewidth]{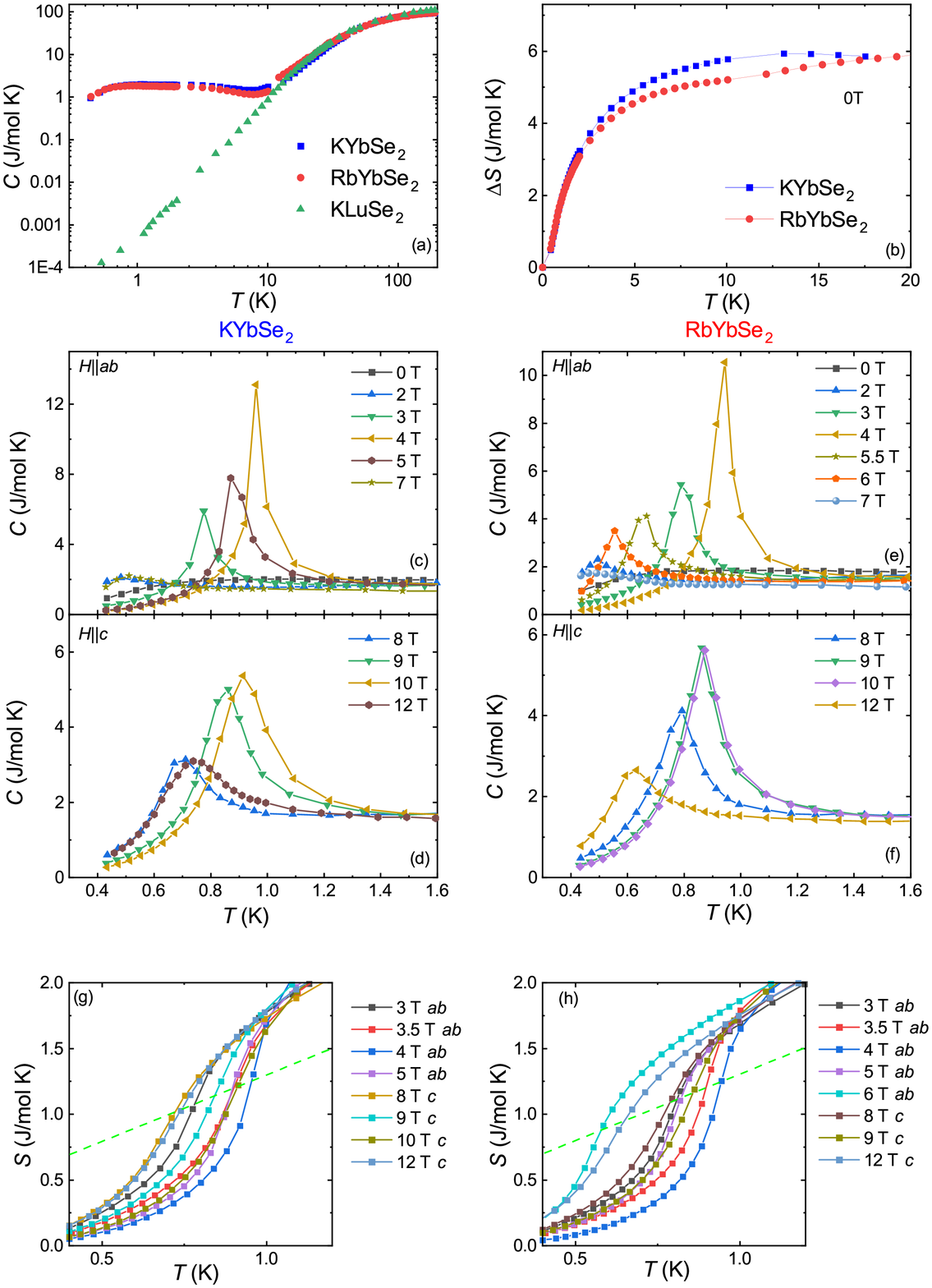}
\caption {\label{fig:hc}(a) Temperature dependence of heat capacity for KYbSe$_2$, RbYbSe$_2$, and KLuSe$_2$ under zero magnetic field above 0.4 K. (b) Temperature dependence of entropy in KYbSe$_2$ and RbYbSe$_2$ at 0 T below 20 K. (c-f) Temperature dependence of heat capacity for KYbSe$_2$ (c,d) and RbYbSe$_2$ (e, f) under different magnetic fields when $H$||$ab$ or $H$||$c$. (g-h) Temperature dependent entropy of KYbSe$_2$ and RbYbSe$_2$ under magnetic fields when the fields are applied in $ab$-plane or $c$-axis. }
\label{crystal}
\end{figure}

Heat capacity has been performed on single crystals down to 0.4 K with $H$||$ab$ and also $H$||$c$. The temperature dependence of heat capacity of K/RbYbSe$_2$ from 0.4 to 200 K at 0 T is shown in Fig.~\ref{fig:hc}(a); key feature is the broad peak from 0.4 to 10 K. This behavior reveals the short-range interactions in the frustrated magnetic layers. No $\lambda$ anomaly is found above 0.4 K, which is consistent with no long-range order in the magnetization at low fields. The heat capacity of isostructural nonmagnetic KLuSe$_2$ single crystal is used as the phonon background. The temperature dependence of entropy in K/RbYbSe$_2$ at 0 T is presented in Fig.~\ref{fig:hc}(b). The magnetic entropy reaches a plateau of $\sim$5.8 J/mol K above 10 K. The slight deviation at 10 K between KYbSe$_2$ and RbYbSe$_2$ may be caused by slightly different interactions between Yb$^{3+}$ or phonon contribution. This value is close to Rln2 of spin 1/2 system, confirming the effective spin 1/2 of Yb$^{3+}$ in this family. 

The heat capacity is also measured when magnetic field is applied along $ab$-plane and $c$-axis for both samples, as shown in Fig.~\ref{fig:hc}(c-f). In these measurements, the field-induced magnetic transition appears and increases to the highest temperature $\sim$ 1 K with the largest peak at the moderate field, then decreases to lower temperature with higher fields. Comparing to the transition at $H$||$ab$, a much higher field is required at $H$||$c$ due to the easy-plane feature. There are long tails above the transition temperature, indicating the strong spin fluctuations from Yb$^{3+}$. The shape of the transition anomaly is comparable to the previous reports in NaYbSe$_2$, and much sharper than in CsYbSe$_2$.~\cite{Ranjith2019naybse,PhysRevB.100.220407} It could be due to the spin correlations in the different space groups of $A$YbSe$_2$, where more 2D character can be found in CsYbSe$_2$ compare to the rest of the compounds in $A$YbSe$_2$ series.

Based on the heat capacity results, we calculate the entropy for all fields in both samples and plot with the applied magnetic fields in Fig.~\ref{fig:hc}(g-h). We use power law fitting to estimate the magnetic entropy between 0 K and 0.4 K, similar to the previous study in NaYbSe$_2$.~\cite{Ranjith2019naybse} The entropy releases up to only $\sim$1.2 J/mol K at the highest transition temperature $\sim$ 1 K, which is close to 1/5 of Rln2 in spin-1/2 system. We find a quasi-linear relation between the field-induced transition temperature and released entropy at transition temperature, marked as green dashed line in Fig.~\ref{fig:hc}(g-h). Above the long-range order temperature, the release entropy merged to $\sim$1.9 J/mol K at $\sim$1.1 K. As the transition temperature varies in a relatively large temperature region with different fields in $A$YbSe$_2$, from $\sim$0.48 K to $\sim$0.96 K, much sharper $\lambda$ anomaly appears and hinting that thermal fluctuation rarely related to the long tails feature. Besides, the released entropy is also independent of magnetic field directions based on the similar entropy vs. temperature behavior of $H$||$ab$ and $H$||$c$ in K/RbYbSe$_2$, although the field-induced spin states are different. We also compare the result of the previously reported CsYbSe$_2$ and find similar relation.~\cite{PhysRevB.100.220407} All the results show the similar relation between the released entropy and temperature under different magnetic fields in this Yb family, indicating two dimensional magnetism in them and weakly related to three dimensional stacking or interlayer interactions.

\begin{figure}[tbh]
\includegraphics[width=1\linewidth]{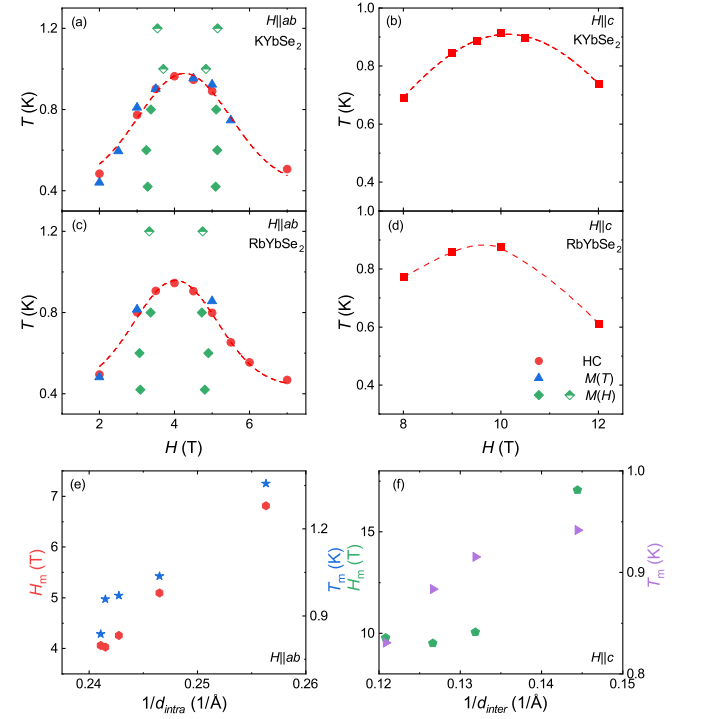}
\caption {\label{fig:pd}(a-d) $H$-$T$ phase diagram of KYbSe$_2$ and RbYbSe$_2$ at $H$||$ab$ and $H$||$c$. The red dashed line presents the possible phase boundary. (e-f) The highest field induced transition temperature and corresponding magnetic field versus inversed intralayer distance and interlayer distance in NaYbS$_2$ and Na/K/Rb/CsYbSe$_2$.}
\label{crystal}
\end{figure}

At last, we make the $H$-$T$ phase diagram of KYbSe$_2$ and RbYbSe$_2$ based on the magnetization and heat capacity results, shown in Fig.~\ref{fig:pd}(a-d). The obtained transition temperatures from the magnetization and heat capacity are consistent in both samples. When the fields are applied in the ab-plane, KYbSe$_2$ presents dome shape long-range order, separated into three different states by the solid green diamonds (the boundaries of the plateau), which can be recognized as 120$^\circ$, collinear up-up-down and 2:1 coplanar spin arrangements. When the magnetic field is applied along c-axis, a similar dome feature is found at a higher field due to the easy-plane feature. The similar phase diagram of RbYbSe$_2$ is shown in Fig.~\ref{fig:pd}(c-d). The half solid diamonds above the long-range order indicate the strong spin fluctuation. The width between the plateau boundaries is almost independent with the temperature, suggesting thermal fluctuation is barely related to the plateau. We could find the domes shift to the lower field from KYbSe$_2$ to RbYbSe$_2$ along both $ab$ and $c$ direction. This is due to the large intralayer and interlayer distance in RbYbSe$_2$.

Considering the previous experimental results of $A$YbSe$_2$ and this present work, the field-induced transitions present clear anisotropic behavior in this family. This could be related to the intralayer and interlayer distance of magnetic Yb$^{3+}$ ions. Based on this assumption, we obtain each highest transition temperature $T_\mathrm{m}$ with the corresponding field $H_\mathrm{m}$ in NaYbS$_2$ and Na/K/Rb/CsYbSe$_2$ series,~\cite{Ranjith2019naybse,PhysRevB.100.220407,ma2020spin} and plotted them against the inversed intralayer distance (1/d$_\mathrm{intra}$) of Yb$^{3+}$ and interlayer distance (1/d$_\mathrm{inter}$), as shown in Fig.~\ref{fig:pd}(e-f). Since the similar $S_{\mathrm{eff}}$=1/2 state of Yb$^{3+}$ is in this family, the interaction energy should directly relate to the magnetic fields and the temperature. Therefore, H$_m$ and T$_m$ present similar behavior against structure parameters in these systems. As displayed in Figure 7(e), both  $H_\mathrm{m}$ and $T_\mathrm{m}$ increase monotonously with 1/d$_\mathrm{intra}$ or 1/d$_\mathrm{inter}$. It is noteworthy to point that, with the smallest value of 1/d$_\mathrm{intra}$ or 1/d$_\mathrm{inter}$ in this family, CsYbSe$_2$ largely deviates from the rest of the structures, Fig~\ref{fig:pd}(e-f). This behavior further implies that intralayer and interlayer interactions between Yb$^{3+}$ triangular magnetic lattice of CsYbSe$_2$ are strongly affected by the size of the Cs$^+$ ion and the packing sequence in the overall 2D structure.

\section{\label{sec:level1}Conclusion}

The KYbSe$_2$ and RbYbSe$_2$ single crystals have been synthesized by flux method and characterized by magnetization and heat capacity. There is no long-range order in both samples above 0.42 K at zero field, revealing the possible spin liquid ground state. Large magnetic anisotropy with easy-plane feature is found in both samples. Magnetic field application can induce the long-range order along both $ab$ and $c$ crystallographic directions, forming similar dome features in $H$-$T$ phase diagram. The temperature dependent magnetic entropy indicates the field-induced magnetic transitions and strong spin fluctuations. The 1/3 magnetic plateau is found in each $A$YbSe$_2$ compound when the magnetic field is applied in the $ab$ plane. At last, we proposed phase diagrams for K/RbYbSe$_2$ single crystals and the empirical relation between field-induced transition and intralayer/interlayer distance. All these results indicate the $A$YbSe$_2$ family has a effective spin-1/2 frustrated magnetism, which is close to QSL and noncolinear spin states.

\begin{acknowledgments}
The research at the Oak Ridge National Laboratory (ORNL) is supported by the U.S. Department of Energy (DOE), Office of Science, Basic Energy Sciences (BES), Materials Sciences and Engineering Division (MSE).
This manuscript has been authored by UT-Battelle, LLC under Contract No. DE-AC05-00OR22725 with the US Department of Energy. The United States Government retains and the publisher, by accepting the article for publication, acknowledges that the United States Government retains a nonexclusive, paid-up, irrevocable, worldwide license to publish or reproduce the published form of this manuscript, or allow others to do so, for United States Government purposes. The Department of Energy will provide public access to these results of federally sponsored research in accordance with the DOE Public Access Plan(http://energy.gov/downloads/doe-public-access-plan).

\end{acknowledgments}

\bibliography{delafossites}

\begin{thebibliography}{65}%
\makeatletter
\providecommand \@ifxundefined [1]{%
 \@ifx{#1\undefined}
}%
\providecommand \@ifnum [1]{%
 \ifnum #1\expandafter \@firstoftwo
 \else \expandafter \@secondoftwo
 \fi
}%
\providecommand \@ifx [1]{%
 \ifx #1\expandafter \@firstoftwo
 \else \expandafter \@secondoftwo
 \fi
}%
\providecommand \natexlab [1]{#1}%
\providecommand \enquote  [1]{``#1''}%
\providecommand \bibnamefont  [1]{#1}%
\providecommand \bibfnamefont [1]{#1}%
\providecommand \citenamefont [1]{#1}%
\providecommand \href@noop [0]{\@secondoftwo}%
\providecommand \href [0]{\begingroup \@sanitize@url \@href}%
\providecommand \@href[1]{\@@startlink{#1}\@@href}%
\providecommand \@@href[1]{\endgroup#1\@@endlink}%
\providecommand \@sanitize@url [0]{\catcode `\\12\catcode `\$12\catcode
  `\&12\catcode `\#12\catcode `\^12\catcode `\_12\catcode `\%12\relax}%
\providecommand \@@startlink[1]{}%
\providecommand \@@endlink[0]{}%
\providecommand \url  [0]{\begingroup\@sanitize@url \@url }%
\providecommand \@url [1]{\endgroup\@href {#1}{\urlprefix }}%
\providecommand \urlprefix  [0]{URL }%
\providecommand \Eprint [0]{\href }%
\providecommand \doibase [0]{http://dx.doi.org/}%
\providecommand \selectlanguage [0]{\@gobble}%
\providecommand \bibinfo  [0]{\@secondoftwo}%
\providecommand \bibfield  [0]{\@secondoftwo}%
\providecommand \translation [1]{[#1]}%
\providecommand \BibitemOpen [0]{}%
\providecommand \bibitemStop [0]{}%
\providecommand \bibitemNoStop [0]{.\EOS\space}%
\providecommand \EOS [0]{\spacefactor3000\relax}%
\providecommand \BibitemShut  [1]{\csname bibitem#1\endcsname}%
\let\auto@bib@innerbib\@empty
\bibitem [{\citenamefont {Sadoc}\ and\ \citenamefont
  {Mosseri}(1999)}]{sadoc_mosseri_1999}%
  \BibitemOpen
  \bibfield  {author} {\bibinfo {author} {\bibfnamefont {J.}~\bibnamefont
  {Sadoc}}\ and\ \bibinfo {author} {\bibfnamefont {R.}~\bibnamefont
  {Mosseri}},\ }\href {\doibase 10.1017/CBO9780511599934} {\emph {\bibinfo
  {title} {Geometrical Frustration}}},\ Collection Alea-Saclay: Monographs and
  Texts in Statistical Physics\ (\bibinfo  {publisher} {Cambridge University
  Press},\ \bibinfo {year} {1999})\BibitemShut {NoStop}%
\bibitem [{\citenamefont {Moessner}\ and\ \citenamefont
  {Ramirez}(2006)}]{Moessner}%
  \BibitemOpen
  \bibfield  {author} {\bibinfo {author} {\bibfnamefont {R.}~\bibnamefont
  {Moessner}}\ and\ \bibinfo {author} {\bibfnamefont {A.~P.}\ \bibnamefont
  {Ramirez}},\ }\bibfield  {title} {\enquote {\bibinfo {title} {Geometrical
  frustration},}\ }\href {\doibase 10.1063/1.2186278} {\bibfield  {journal}
  {\bibinfo  {journal} {Phys. Today}\ }\textbf {\bibinfo {volume} {59}},\
  \bibinfo {pages} {24--29} (\bibinfo {year} {2006})}\BibitemShut {NoStop}%
\bibitem [{\citenamefont {Anderson}(1973)}]{Anderson}%
  \BibitemOpen
  \bibfield  {author} {\bibinfo {author} {\bibfnamefont {P.}~\bibnamefont
  {Anderson}},\ }\bibfield  {title} {\enquote {\bibinfo {title} {{Resonating
  valence bonds: A new kind of insulator?}}}\ }\href {\doibase
  https://doi.org/10.1016/0025-5408(73)90167-0} {\bibfield  {journal} {\bibinfo
   {journal} {Mater. Res. Bull.}\ }\textbf {\bibinfo {volume} {8}},\ \bibinfo
  {pages} {153 -- 160} (\bibinfo {year} {1973})}\BibitemShut {NoStop}%
\bibitem [{\citenamefont {Mila}(2000)}]{Mila2000}%
  \BibitemOpen
  \bibfield  {author} {\bibinfo {author} {\bibfnamefont {F.}~\bibnamefont
  {Mila}},\ }\bibfield  {title} {\enquote {\bibinfo {title} {Quantum spin
  liquids},}\ }\href {\doibase 10.1088/0143-0807/21/6/302} {\bibfield
  {journal} {\bibinfo  {journal} {Eur. J. Phys.}\ }\textbf {\bibinfo {volume}
  {21}},\ \bibinfo {pages} {499--510} (\bibinfo {year} {2000})}\BibitemShut
  {NoStop}%
\bibitem [{\citenamefont {Balents}(2010)}]{Balents}%
  \BibitemOpen
  \bibfield  {author} {\bibinfo {author} {\bibfnamefont {L.}~\bibnamefont
  {Balents}},\ }\bibfield  {title} {\enquote {\bibinfo {title} {Spin liquids in
  frustrated magnets},}\ }\href {\doibase https://doi.org/10.1038/nature08917}
  {\bibfield  {journal} {\bibinfo  {journal} {Nature}\ }\textbf {\bibinfo
  {volume} {464}},\ \bibinfo {pages} {199--208} (\bibinfo {year}
  {2010})}\BibitemShut {NoStop}%
\bibitem [{\citenamefont {Gardner}, \citenamefont {Gingras},\ and\
  \citenamefont {Greedan}(2010)}]{RevModPhys}%
  \BibitemOpen
  \bibfield  {author} {\bibinfo {author} {\bibfnamefont {J.~S.}\ \bibnamefont
  {Gardner}}, \bibinfo {author} {\bibfnamefont {M.~J.~P.}\ \bibnamefont
  {Gingras}}, \ and\ \bibinfo {author} {\bibfnamefont {J.~E.}\ \bibnamefont
  {Greedan}},\ }\bibfield  {title} {\enquote {\bibinfo {title} {{Magnetic
  pyrochlore oxides}},}\ }\href {\doibase 10.1103/RevModPhys.82.53} {\bibfield
  {journal} {\bibinfo  {journal} {Rev. Mod. Phys.}\ }\textbf {\bibinfo {volume}
  {82}},\ \bibinfo {pages} {53--107} (\bibinfo {year} {2010})}\BibitemShut
  {NoStop}%
\bibitem [{\citenamefont {Broholm}\ \emph {et~al.}(2020)\citenamefont
  {Broholm}, \citenamefont {Cava}, \citenamefont {Kivelson}, \citenamefont
  {Nocera}, \citenamefont {Norman},\ and\ \citenamefont
  {Senthil}}]{Broholmeaay0668}%
  \BibitemOpen
  \bibfield  {author} {\bibinfo {author} {\bibfnamefont {C.}~\bibnamefont
  {Broholm}}, \bibinfo {author} {\bibfnamefont {R.~J.}\ \bibnamefont {Cava}},
  \bibinfo {author} {\bibfnamefont {S.~A.}\ \bibnamefont {Kivelson}}, \bibinfo
  {author} {\bibfnamefont {D.~G.}\ \bibnamefont {Nocera}}, \bibinfo {author}
  {\bibfnamefont {M.~R.}\ \bibnamefont {Norman}}, \ and\ \bibinfo {author}
  {\bibfnamefont {T.}~\bibnamefont {Senthil}},\ }\bibfield  {title} {\enquote
  {\bibinfo {title} {Quantum spin liquids},}\ }\href {\doibase
  10.1126/science.aay0668} {\bibfield  {journal} {\bibinfo  {journal}
  {Science}\ }\textbf {\bibinfo {volume} {367}},\ \bibinfo {pages} {6475}
  (\bibinfo {year} {2020})}\BibitemShut {NoStop}%
\bibitem [{\citenamefont {Li}, \citenamefont {Gegenwart},\ and\ \citenamefont
  {Tsirlin}(2020)}]{li2020spin}%
  \BibitemOpen
  \bibfield  {author} {\bibinfo {author} {\bibfnamefont {Y.}~\bibnamefont
  {Li}}, \bibinfo {author} {\bibfnamefont {P.}~\bibnamefont {Gegenwart}}, \
  and\ \bibinfo {author} {\bibfnamefont {A.~A.}\ \bibnamefont {Tsirlin}},\
  }\bibfield  {title} {\enquote {\bibinfo {title} {{Spin liquids in
  geometrically perfect triangular antiferromagnets}},}\ }\href@noop {}
  {\bibfield  {journal} {\bibinfo  {journal} {Journal of Physics: Condensed
  Matter}\ }\textbf {\bibinfo {volume} {32}},\ \bibinfo {pages} {224004}
  (\bibinfo {year} {2020})}\BibitemShut {NoStop}%
\bibitem [{\citenamefont {Sarkis}\ \emph {et~al.}(2020)\citenamefont {Sarkis},
  \citenamefont {Rau}, \citenamefont {Sanjeewa}, \citenamefont {Powell},
  \citenamefont {Kolis}, \citenamefont {Marbey}, \citenamefont {Hill},
  \citenamefont {Rodriguez-Rivera}, \citenamefont {Nair}, \citenamefont {Yahne}
  \emph {et~al.}}]{sarkis2020unravelling}%
  \BibitemOpen
  \bibfield  {author} {\bibinfo {author} {\bibfnamefont {C.}~\bibnamefont
  {Sarkis}}, \bibinfo {author} {\bibfnamefont {J.~G.}\ \bibnamefont {Rau}},
  \bibinfo {author} {\bibfnamefont {L.}~\bibnamefont {Sanjeewa}}, \bibinfo
  {author} {\bibfnamefont {M.}~\bibnamefont {Powell}}, \bibinfo {author}
  {\bibfnamefont {J.}~\bibnamefont {Kolis}}, \bibinfo {author} {\bibfnamefont
  {J.}~\bibnamefont {Marbey}}, \bibinfo {author} {\bibfnamefont
  {S.}~\bibnamefont {Hill}}, \bibinfo {author} {\bibfnamefont {J.}~\bibnamefont
  {Rodriguez-Rivera}}, \bibinfo {author} {\bibfnamefont {H.}~\bibnamefont
  {Nair}}, \bibinfo {author} {\bibfnamefont {D.}~\bibnamefont {Yahne}},  \emph
  {et~al.},\ }\bibfield  {title} {\enquote {\bibinfo {title} {{Unravelling
  competing microscopic interactions at a phase boundary: A single-crystal
  study of the metastable antiferromagnetic pyrochlore Yb$_2$Ge$_2$O$_7$}},}\
  }\href@noop {} {\bibfield  {journal} {\bibinfo  {journal} {Physical Review
  B}\ }\textbf {\bibinfo {volume} {102}},\ \bibinfo {pages} {134418} (\bibinfo
  {year} {2020})}\BibitemShut {NoStop}%
\bibitem [{\citenamefont {Pajerowski}\ \emph {et~al.}(2020)\citenamefont
  {Pajerowski}, \citenamefont {Taddei}, \citenamefont {Sanjeewa}, \citenamefont
  {Savici}, \citenamefont {Stone},\ and\ \citenamefont
  {Kolis}}]{pajerowski2020quantification}%
  \BibitemOpen
  \bibfield  {author} {\bibinfo {author} {\bibfnamefont {D.~M.}\ \bibnamefont
  {Pajerowski}}, \bibinfo {author} {\bibfnamefont {K.~M.}\ \bibnamefont
  {Taddei}}, \bibinfo {author} {\bibfnamefont {L.~D.}\ \bibnamefont
  {Sanjeewa}}, \bibinfo {author} {\bibfnamefont {A.~T.}\ \bibnamefont
  {Savici}}, \bibinfo {author} {\bibfnamefont {M.~B.}\ \bibnamefont {Stone}}, \
  and\ \bibinfo {author} {\bibfnamefont {J.~W.}\ \bibnamefont {Kolis}},\
  }\bibfield  {title} {\enquote {\bibinfo {title} {{Quantification of local
  Ising Magnetism in Rare-earth Pyrogermanates Er$_2$Ge$_2$O$_7$ and
  Yb$_2$Ge$_2$O$_7$}},}\ }\href@noop {} {\bibfield  {journal} {\bibinfo
  {journal} {Physical Review B}\ }\textbf {\bibinfo {volume} {101}},\ \bibinfo
  {pages} {014420} (\bibinfo {year} {2020})}\BibitemShut {NoStop}%
\bibitem [{\citenamefont {Rau}\ and\ \citenamefont
  {Gingras}(2018)}]{rau2018frustration}%
  \BibitemOpen
  \bibfield  {author} {\bibinfo {author} {\bibfnamefont {J.~G.}\ \bibnamefont
  {Rau}}\ and\ \bibinfo {author} {\bibfnamefont {M.~J.}\ \bibnamefont
  {Gingras}},\ }\bibfield  {title} {\enquote {\bibinfo {title} {{Frustration
  and anisotropic exchange in ytterbium magnets with edge-shared octahedra}},}\
  }\href@noop {} {\bibfield  {journal} {\bibinfo  {journal} {Physical Review
  B}\ }\textbf {\bibinfo {volume} {98}},\ \bibinfo {pages} {054408} (\bibinfo
  {year} {2018})}\BibitemShut {NoStop}%
\bibitem [{\citenamefont {Xing}\ \emph {et~al.}(2020)\citenamefont {Xing},
  \citenamefont {Feng}, \citenamefont {Liu}, \citenamefont {Emmanouilidou},
  \citenamefont {Hu}, \citenamefont {Liu}, \citenamefont {Graf}, \citenamefont
  {Ramirez}, \citenamefont {Chen}, \citenamefont {Cao},\ and\ \citenamefont
  {Ni}}]{xingYbCl3}%
  \BibitemOpen
  \bibfield  {author} {\bibinfo {author} {\bibfnamefont {J.}~\bibnamefont
  {Xing}}, \bibinfo {author} {\bibfnamefont {E.}~\bibnamefont {Feng}}, \bibinfo
  {author} {\bibfnamefont {Y.}~\bibnamefont {Liu}}, \bibinfo {author}
  {\bibfnamefont {E.}~\bibnamefont {Emmanouilidou}}, \bibinfo {author}
  {\bibfnamefont {C.}~\bibnamefont {Hu}}, \bibinfo {author} {\bibfnamefont
  {J.}~\bibnamefont {Liu}}, \bibinfo {author} {\bibfnamefont {D.}~\bibnamefont
  {Graf}}, \bibinfo {author} {\bibfnamefont {A.~P.}\ \bibnamefont {Ramirez}},
  \bibinfo {author} {\bibfnamefont {G.}~\bibnamefont {Chen}}, \bibinfo {author}
  {\bibfnamefont {H.}~\bibnamefont {Cao}}, \ and\ \bibinfo {author}
  {\bibfnamefont {N.}~\bibnamefont {Ni}},\ }\bibfield  {title} {\enquote
  {\bibinfo {title} {{N\'eel-type antiferromagnetic order and magnetic
  field--temperature phase diagram in the spin-$\frac{1}{2}$ rare-earth
  honeycomb compound $\mathrm{YbCl}_{3}$}},}\ }\href@noop {} {\bibfield
  {journal} {\bibinfo  {journal} {Phys. Rev. B}\ }\textbf {\bibinfo {volume}
  {102}},\ \bibinfo {pages} {014427} (\bibinfo {year} {2020})}\BibitemShut
  {NoStop}%
\bibitem [{\citenamefont {Wu}\ \emph {et~al.}(2019{\natexlab{a}})\citenamefont
  {Wu}, \citenamefont {Nikitin}, \citenamefont {Wang}, \citenamefont {Zhu},
  \citenamefont {Batista}, \citenamefont {Tsvelik}, \citenamefont {Samarakoon},
  \citenamefont {Tennant}, \citenamefont {Brando}, \citenamefont {Vasylechko},
  \citenamefont {Frontzek}, \citenamefont {Savici}, \citenamefont {Sala},
  \citenamefont {Ehlers}, \citenamefont {Christianson}, \citenamefont
  {Lumsden},\ and\ \citenamefont {Podlesnyak}}]{Wu1}%
  \BibitemOpen
  \bibfield  {author} {\bibinfo {author} {\bibfnamefont {L.~S.}\ \bibnamefont
  {Wu}}, \bibinfo {author} {\bibfnamefont {S.~E.}\ \bibnamefont {Nikitin}},
  \bibinfo {author} {\bibfnamefont {Z.}~\bibnamefont {Wang}}, \bibinfo {author}
  {\bibfnamefont {W.}~\bibnamefont {Zhu}}, \bibinfo {author} {\bibfnamefont
  {C.~D.}\ \bibnamefont {Batista}}, \bibinfo {author} {\bibfnamefont {A.~M.}\
  \bibnamefont {Tsvelik}}, \bibinfo {author} {\bibfnamefont {A.~M.}\
  \bibnamefont {Samarakoon}}, \bibinfo {author} {\bibfnamefont {D.~A.}\
  \bibnamefont {Tennant}}, \bibinfo {author} {\bibfnamefont {M.}~\bibnamefont
  {Brando}}, \bibinfo {author} {\bibfnamefont {L.}~\bibnamefont {Vasylechko}},
  \bibinfo {author} {\bibfnamefont {M.}~\bibnamefont {Frontzek}}, \bibinfo
  {author} {\bibfnamefont {A.~T.}\ \bibnamefont {Savici}}, \bibinfo {author}
  {\bibfnamefont {G.}~\bibnamefont {Sala}}, \bibinfo {author} {\bibfnamefont
  {G.}~\bibnamefont {Ehlers}}, \bibinfo {author} {\bibfnamefont {A.~D.}\
  \bibnamefont {Christianson}}, \bibinfo {author} {\bibfnamefont {M.~D.}\
  \bibnamefont {Lumsden}}, \ and\ \bibinfo {author} {\bibfnamefont
  {A.}~\bibnamefont {Podlesnyak}},\ }\bibfield  {title} {\enquote {\bibinfo
  {title} {{Tomonaga-Luttinger liquid behavior and spinon confinement in
  YbAlO$_3$}},}\ }\href {\doibase 10.1038/s41467-019-08485-7} {\bibfield
  {journal} {\bibinfo  {journal} {Nat. Commun.}\ }\textbf {\bibinfo {volume}
  {10}},\ \bibinfo {pages} {698} (\bibinfo {year}
  {2019}{\natexlab{a}})}\BibitemShut {NoStop}%
\bibitem [{\citenamefont {Wu}\ \emph {et~al.}(2019{\natexlab{b}})\citenamefont
  {Wu}, \citenamefont {Nikitin}, \citenamefont {Brando}, \citenamefont
  {Vasylechko}, \citenamefont {Ehlers}, \citenamefont {Frontzek}, \citenamefont
  {Savici}, \citenamefont {Sala}, \citenamefont {Christianson}, \citenamefont
  {Lumsden},\ and\ \citenamefont {Podlesnyak}}]{Wu2}%
  \BibitemOpen
  \bibfield  {author} {\bibinfo {author} {\bibfnamefont {L.~S.}\ \bibnamefont
  {Wu}}, \bibinfo {author} {\bibfnamefont {S.~E.}\ \bibnamefont {Nikitin}},
  \bibinfo {author} {\bibfnamefont {M.}~\bibnamefont {Brando}}, \bibinfo
  {author} {\bibfnamefont {L.}~\bibnamefont {Vasylechko}}, \bibinfo {author}
  {\bibfnamefont {G.}~\bibnamefont {Ehlers}}, \bibinfo {author} {\bibfnamefont
  {M.}~\bibnamefont {Frontzek}}, \bibinfo {author} {\bibfnamefont {A.~T.}\
  \bibnamefont {Savici}}, \bibinfo {author} {\bibfnamefont {G.}~\bibnamefont
  {Sala}}, \bibinfo {author} {\bibfnamefont {A.~D.}\ \bibnamefont
  {Christianson}}, \bibinfo {author} {\bibfnamefont {M.~D.}\ \bibnamefont
  {Lumsden}}, \ and\ \bibinfo {author} {\bibfnamefont {A.}~\bibnamefont
  {Podlesnyak}},\ }\bibfield  {title} {\enquote {\bibinfo {title}
  {{Antiferromagnetic ordering and dipolar interactions of YbAlO$_3$}},}\
  }\href {\doibase 10.1103/PhysRevB.99.195117} {\bibfield  {journal} {\bibinfo
  {journal} {Phys. Rev. B}\ }\textbf {\bibinfo {volume} {99}},\ \bibinfo
  {pages} {195117} (\bibinfo {year} {2019}{\natexlab{b}})}\BibitemShut
  {NoStop}%
\bibitem [{\citenamefont {Agrapidis}, \citenamefont {van~den Brink},\ and\
  \citenamefont {Nishimoto}(2019)}]{agrapidis}%
  \BibitemOpen
  \bibfield  {author} {\bibinfo {author} {\bibfnamefont {C.~E.}\ \bibnamefont
  {Agrapidis}}, \bibinfo {author} {\bibfnamefont {J.}~\bibnamefont {van~den
  Brink}}, \ and\ \bibinfo {author} {\bibfnamefont {S.}~\bibnamefont
  {Nishimoto}},\ }\bibfield  {title} {\enquote {\bibinfo {title}
  {{Field-induced incommensurate ordering in Heisenberg chains coupled by Ising
  interaction: Model for ytterbium aluminum perovskite YbAlO$_3$}},}\ }\href
  {\doibase 10.1103/PhysRevB.99.224423} {\bibfield  {journal} {\bibinfo
  {journal} {Phys. Rev. B}\ }\textbf {\bibinfo {volume} {99}},\ \bibinfo
  {pages} {224423} (\bibinfo {year} {2019})}\BibitemShut {NoStop}%
\bibitem [{\citenamefont {Hester}\ \emph {et~al.}(2019)\citenamefont {Hester},
  \citenamefont {Nair}, \citenamefont {Reeder}, \citenamefont {Yahne},
  \citenamefont {DeLazzer}, \citenamefont {Berges}, \citenamefont {Ziat},
  \citenamefont {Neilson}, \citenamefont {Aczel}, \citenamefont {Sala},
  \citenamefont {Quilliam},\ and\ \citenamefont {Ross}}]{Hester}%
  \BibitemOpen
  \bibfield  {author} {\bibinfo {author} {\bibfnamefont {G.}~\bibnamefont
  {Hester}}, \bibinfo {author} {\bibfnamefont {H.~S.}\ \bibnamefont {Nair}},
  \bibinfo {author} {\bibfnamefont {T.}~\bibnamefont {Reeder}}, \bibinfo
  {author} {\bibfnamefont {D.~R.}\ \bibnamefont {Yahne}}, \bibinfo {author}
  {\bibfnamefont {T.~N.}\ \bibnamefont {DeLazzer}}, \bibinfo {author}
  {\bibfnamefont {L.}~\bibnamefont {Berges}}, \bibinfo {author} {\bibfnamefont
  {D.}~\bibnamefont {Ziat}}, \bibinfo {author} {\bibfnamefont {J.~R.}\
  \bibnamefont {Neilson}}, \bibinfo {author} {\bibfnamefont {A.~A.}\
  \bibnamefont {Aczel}}, \bibinfo {author} {\bibfnamefont {G.}~\bibnamefont
  {Sala}}, \bibinfo {author} {\bibfnamefont {J.~A.}\ \bibnamefont {Quilliam}},
  \ and\ \bibinfo {author} {\bibfnamefont {K.~A.}\ \bibnamefont {Ross}},\
  }\bibfield  {title} {\enquote {\bibinfo {title} {{Novel Strongly Spin-Orbit
  Coupled Quantum Dimer Magnet: Yb$_2$Si$_2$O$_7$}},}\ }\href {\doibase
  10.1103/PhysRevLett.123.027201} {\bibfield  {journal} {\bibinfo  {journal}
  {Phys. Rev. Lett.}\ }\textbf {\bibinfo {volume} {123}},\ \bibinfo {pages}
  {027201} (\bibinfo {year} {2019})}\BibitemShut {NoStop}%
\bibitem [{\citenamefont {Li}, \citenamefont {Lu},\ and\ \citenamefont
  {Chen}(2017)}]{lispinon}%
  \BibitemOpen
  \bibfield  {author} {\bibinfo {author} {\bibfnamefont {Y.-D.}\ \bibnamefont
  {Li}}, \bibinfo {author} {\bibfnamefont {Y.-M.}\ \bibnamefont {Lu}}, \ and\
  \bibinfo {author} {\bibfnamefont {G.}~\bibnamefont {Chen}},\ }\bibfield
  {title} {\enquote {\bibinfo {title} {Spinon fermi surface $u(1)$ spin liquid
  in the spin-orbit-coupled triangular-lattice mott insulator
  ${\mathrm{ybmggao}}_{4}$},}\ }\href {\doibase 10.1103/PhysRevB.96.054445}
  {\bibfield  {journal} {\bibinfo  {journal} {Phys. Rev. B}\ }\textbf {\bibinfo
  {volume} {96}},\ \bibinfo {pages} {054445} (\bibinfo {year}
  {2017})}\BibitemShut {NoStop}%
\bibitem [{\citenamefont {Li}\ \emph {et~al.}(2018)\citenamefont {Li},
  \citenamefont {Shen}, \citenamefont {Li}, \citenamefont {Zhao},\ and\
  \citenamefont {Chen}}]{lieffect}%
  \BibitemOpen
  \bibfield  {author} {\bibinfo {author} {\bibfnamefont {Y.-D.}\ \bibnamefont
  {Li}}, \bibinfo {author} {\bibfnamefont {Y.}~\bibnamefont {Shen}}, \bibinfo
  {author} {\bibfnamefont {Y.}~\bibnamefont {Li}}, \bibinfo {author}
  {\bibfnamefont {J.}~\bibnamefont {Zhao}}, \ and\ \bibinfo {author}
  {\bibfnamefont {G.}~\bibnamefont {Chen}},\ }\bibfield  {title} {\enquote
  {\bibinfo {title} {Effect of spin-orbit coupling on the effective-spin
  correlation in ${\mathrm{ybmggao}}_{4}$},}\ }\href {\doibase
  10.1103/PhysRevB.97.125105} {\bibfield  {journal} {\bibinfo  {journal} {Phys.
  Rev. B}\ }\textbf {\bibinfo {volume} {97}},\ \bibinfo {pages} {125105}
  (\bibinfo {year} {2018})}\BibitemShut {NoStop}%
\bibitem [{\citenamefont {Li}, \citenamefont {Wang},\ and\ \citenamefont
  {Chen}(2016)}]{lianisotropic}%
  \BibitemOpen
  \bibfield  {author} {\bibinfo {author} {\bibfnamefont {Y.-D.}\ \bibnamefont
  {Li}}, \bibinfo {author} {\bibfnamefont {X.}~\bibnamefont {Wang}}, \ and\
  \bibinfo {author} {\bibfnamefont {G.}~\bibnamefont {Chen}},\ }\bibfield
  {title} {\enquote {\bibinfo {title} {Anisotropic spin model of strong
  spin-orbit-coupled triangular antiferromagnets},}\ }\href {\doibase
  10.1103/PhysRevB.94.035107} {\bibfield  {journal} {\bibinfo  {journal} {Phys.
  Rev. B}\ }\textbf {\bibinfo {volume} {94}},\ \bibinfo {pages} {035107}
  (\bibinfo {year} {2016})}\BibitemShut {NoStop}%
\bibitem [{\citenamefont {Li}\ \emph {et~al.}(2015{\natexlab{a}})\citenamefont
  {Li}, \citenamefont {Liao}, \citenamefont {Zhang}, \citenamefont {Li},
  \citenamefont {Jin}, \citenamefont {Ling}, \citenamefont {Zhang},
  \citenamefont {Zou}, \citenamefont {Pi}, \citenamefont {Yang} \emph
  {et~al.}}]{li2015gapless}%
  \BibitemOpen
  \bibfield  {author} {\bibinfo {author} {\bibfnamefont {Y.}~\bibnamefont
  {Li}}, \bibinfo {author} {\bibfnamefont {H.}~\bibnamefont {Liao}}, \bibinfo
  {author} {\bibfnamefont {Z.}~\bibnamefont {Zhang}}, \bibinfo {author}
  {\bibfnamefont {S.}~\bibnamefont {Li}}, \bibinfo {author} {\bibfnamefont
  {F.}~\bibnamefont {Jin}}, \bibinfo {author} {\bibfnamefont {L.}~\bibnamefont
  {Ling}}, \bibinfo {author} {\bibfnamefont {L.}~\bibnamefont {Zhang}},
  \bibinfo {author} {\bibfnamefont {Y.}~\bibnamefont {Zou}}, \bibinfo {author}
  {\bibfnamefont {L.}~\bibnamefont {Pi}}, \bibinfo {author} {\bibfnamefont
  {Z.}~\bibnamefont {Yang}},  \emph {et~al.},\ }\bibfield  {title} {\enquote
  {\bibinfo {title} {{Gapless quantum spin liquid ground state in the
  two-dimensional spin-1/2 triangular antiferromagnet YbMgGaO$_4$}},}\ }\href
  {\doibase doi: 10.1038/srep16419} {\bibfield  {journal} {\bibinfo  {journal}
  {Scientific reports}\ }\textbf {\bibinfo {volume} {5}},\ \bibinfo {pages}
  {16419} (\bibinfo {year} {2015}{\natexlab{a}})}\BibitemShut {NoStop}%
\bibitem [{\citenamefont {Li}\ \emph {et~al.}(2015{\natexlab{b}})\citenamefont
  {Li}, \citenamefont {Chen}, \citenamefont {Tong}, \citenamefont {Pi},
  \citenamefont {Liu}, \citenamefont {Yang}, \citenamefont {Wang},\ and\
  \citenamefont {Zhang}}]{li2015rare}%
  \BibitemOpen
  \bibfield  {author} {\bibinfo {author} {\bibfnamefont {Y.}~\bibnamefont
  {Li}}, \bibinfo {author} {\bibfnamefont {G.}~\bibnamefont {Chen}}, \bibinfo
  {author} {\bibfnamefont {W.}~\bibnamefont {Tong}}, \bibinfo {author}
  {\bibfnamefont {L.}~\bibnamefont {Pi}}, \bibinfo {author} {\bibfnamefont
  {J.}~\bibnamefont {Liu}}, \bibinfo {author} {\bibfnamefont {Z.}~\bibnamefont
  {Yang}}, \bibinfo {author} {\bibfnamefont {X.}~\bibnamefont {Wang}}, \ and\
  \bibinfo {author} {\bibfnamefont {Q.}~\bibnamefont {Zhang}},\ }\bibfield
  {title} {\enquote {\bibinfo {title} {{Rare-Earth Triangular Lattice Spin
  Liquid: A Single-Crystal Study of ${\mathrm{YbMgGaO}}_{4}$}},}\ }\href
  {\doibase 10.1103/PhysRevLett.115.167203} {\bibfield  {journal} {\bibinfo
  {journal} {Phys. Rev. Lett.}\ }\textbf {\bibinfo {volume} {115}},\ \bibinfo
  {pages} {167203} (\bibinfo {year} {2015}{\natexlab{b}})}\BibitemShut
  {NoStop}%
\bibitem [{\citenamefont {Li}\ \emph {et~al.}(2016)\citenamefont {Li},
  \citenamefont {Adroja}, \citenamefont {Biswas}, \citenamefont {Baker},
  \citenamefont {Zhang}, \citenamefont {Liu}, \citenamefont {Tsirlin},
  \citenamefont {Gegenwart},\ and\ \citenamefont {Zhang}}]{li2016muon}%
  \BibitemOpen
  \bibfield  {author} {\bibinfo {author} {\bibfnamefont {Y.}~\bibnamefont
  {Li}}, \bibinfo {author} {\bibfnamefont {D.}~\bibnamefont {Adroja}}, \bibinfo
  {author} {\bibfnamefont {P.~K.}\ \bibnamefont {Biswas}}, \bibinfo {author}
  {\bibfnamefont {P.~J.}\ \bibnamefont {Baker}}, \bibinfo {author}
  {\bibfnamefont {Q.}~\bibnamefont {Zhang}}, \bibinfo {author} {\bibfnamefont
  {J.}~\bibnamefont {Liu}}, \bibinfo {author} {\bibfnamefont {A.~A.}\
  \bibnamefont {Tsirlin}}, \bibinfo {author} {\bibfnamefont {P.}~\bibnamefont
  {Gegenwart}}, \ and\ \bibinfo {author} {\bibfnamefont {Q.}~\bibnamefont
  {Zhang}},\ }\bibfield  {title} {\enquote {\bibinfo {title} {{Muon Spin
  Relaxation Evidence for the U(1) Quantum Spin-Liquid Ground State in the
  Triangular Antiferromagnet ${\mathrm{YbMgGaO}}_{4}$}},}\ }\href {\doibase
  10.1103/PhysRevLett.117.097201} {\bibfield  {journal} {\bibinfo  {journal}
  {Phys. Rev. Lett.}\ }\textbf {\bibinfo {volume} {117}},\ \bibinfo {pages}
  {097201} (\bibinfo {year} {2016})}\BibitemShut {NoStop}%
\bibitem [{\citenamefont {Shen}\ \emph {et~al.}(2016)\citenamefont {Shen},
  \citenamefont {Li}, \citenamefont {Wo}, \citenamefont {Li}, \citenamefont
  {Shen}, \citenamefont {Pan}, \citenamefont {Wang}, \citenamefont {Walker},
  \citenamefont {Steffens}, \citenamefont {Boehm} \emph
  {et~al.}}]{shen2016evidence}%
  \BibitemOpen
  \bibfield  {author} {\bibinfo {author} {\bibfnamefont {Y.}~\bibnamefont
  {Shen}}, \bibinfo {author} {\bibfnamefont {Y.-D.}\ \bibnamefont {Li}},
  \bibinfo {author} {\bibfnamefont {H.}~\bibnamefont {Wo}}, \bibinfo {author}
  {\bibfnamefont {Y.}~\bibnamefont {Li}}, \bibinfo {author} {\bibfnamefont
  {S.}~\bibnamefont {Shen}}, \bibinfo {author} {\bibfnamefont {B.}~\bibnamefont
  {Pan}}, \bibinfo {author} {\bibfnamefont {Q.}~\bibnamefont {Wang}}, \bibinfo
  {author} {\bibfnamefont {H.}~\bibnamefont {Walker}}, \bibinfo {author}
  {\bibfnamefont {P.}~\bibnamefont {Steffens}}, \bibinfo {author}
  {\bibfnamefont {M.}~\bibnamefont {Boehm}},  \emph {et~al.},\ }\bibfield
  {title} {\enquote {\bibinfo {title} {{Evidence for a spinon Fermi surface in
  a triangular-lattice quantum-spin-liquid candidate}},}\ }\href {\doibase
  https://doi.org/10.1038/nature20614} {\bibfield  {journal} {\bibinfo
  {journal} {Nature}\ }\textbf {\bibinfo {volume} {540}},\ \bibinfo {pages}
  {559} (\bibinfo {year} {2016})}\BibitemShut {NoStop}%
\bibitem [{\citenamefont {Paddison}\ \emph {et~al.}(2017)\citenamefont
  {Paddison}, \citenamefont {Daum}, \citenamefont {Dun}, \citenamefont
  {Ehlers}, \citenamefont {Liu}, \citenamefont {Stone}, \citenamefont {Zhou},\
  and\ \citenamefont {Mourigal}}]{paddison2017continuous}%
  \BibitemOpen
  \bibfield  {author} {\bibinfo {author} {\bibfnamefont {J.~A.~M.}\
  \bibnamefont {Paddison}}, \bibinfo {author} {\bibfnamefont {M.}~\bibnamefont
  {Daum}}, \bibinfo {author} {\bibfnamefont {Z.}~\bibnamefont {Dun}}, \bibinfo
  {author} {\bibfnamefont {G.}~\bibnamefont {Ehlers}}, \bibinfo {author}
  {\bibfnamefont {Y.}~\bibnamefont {Liu}}, \bibinfo {author} {\bibfnamefont
  {M.~B.}\ \bibnamefont {Stone}}, \bibinfo {author} {\bibfnamefont
  {H.}~\bibnamefont {Zhou}}, \ and\ \bibinfo {author} {\bibfnamefont
  {M.}~\bibnamefont {Mourigal}},\ }\bibfield  {title} {\enquote {\bibinfo
  {title} {{Continuous excitations of the triangular-lattice quantum spin
  liquid YbMgGaO$_4$}},}\ }\href {https://www.nature.com/articles/nphys3971}
  {\bibfield  {journal} {\bibinfo  {journal} {Nat. Phys.}\ }\textbf {\bibinfo
  {volume} {13}},\ \bibinfo {pages} {117} (\bibinfo {year} {2017})}\BibitemShut
  {NoStop}%
\bibitem [{\citenamefont {Zhang}\ \emph {et~al.}(2018)\citenamefont {Zhang},
  \citenamefont {Mahmood}, \citenamefont {Daum}, \citenamefont {Dun},
  \citenamefont {Paddison}, \citenamefont {Laurita}, \citenamefont {Hong},
  \citenamefont {Zhou}, \citenamefont {Armitage},\ and\ \citenamefont
  {Mourigal}}]{zhang2018hierarchy}%
  \BibitemOpen
  \bibfield  {author} {\bibinfo {author} {\bibfnamefont {X.}~\bibnamefont
  {Zhang}}, \bibinfo {author} {\bibfnamefont {F.}~\bibnamefont {Mahmood}},
  \bibinfo {author} {\bibfnamefont {M.}~\bibnamefont {Daum}}, \bibinfo {author}
  {\bibfnamefont {Z.}~\bibnamefont {Dun}}, \bibinfo {author} {\bibfnamefont
  {J.~A.~M.}\ \bibnamefont {Paddison}}, \bibinfo {author} {\bibfnamefont
  {N.~J.}\ \bibnamefont {Laurita}}, \bibinfo {author} {\bibfnamefont
  {T.}~\bibnamefont {Hong}}, \bibinfo {author} {\bibfnamefont {H.}~\bibnamefont
  {Zhou}}, \bibinfo {author} {\bibfnamefont {N.~P.}\ \bibnamefont {Armitage}},
  \ and\ \bibinfo {author} {\bibfnamefont {M.}~\bibnamefont {Mourigal}},\
  }\bibfield  {title} {\enquote {\bibinfo {title} {{Hierarchy of Exchange
  Interactions in the Triangular-Lattice Spin Liquid
  ${\mathrm{YbMgGaO}}_{4}$}},}\ }\href {\doibase 10.1103/PhysRevX.8.031001}
  {\bibfield  {journal} {\bibinfo  {journal} {Phys. Rev. X}\ }\textbf {\bibinfo
  {volume} {8}},\ \bibinfo {pages} {031001} (\bibinfo {year}
  {2018})}\BibitemShut {NoStop}%
\bibitem [{\citenamefont {Li}\ \emph {et~al.}(2017)\citenamefont {Li},
  \citenamefont {Adroja}, \citenamefont {Bewley}, \citenamefont {Voneshen},
  \citenamefont {Tsirlin}, \citenamefont {Gegenwart},\ and\ \citenamefont
  {Zhang}}]{li2017crystalline}%
  \BibitemOpen
  \bibfield  {author} {\bibinfo {author} {\bibfnamefont {Y.}~\bibnamefont
  {Li}}, \bibinfo {author} {\bibfnamefont {D.}~\bibnamefont {Adroja}}, \bibinfo
  {author} {\bibfnamefont {R.~I.}\ \bibnamefont {Bewley}}, \bibinfo {author}
  {\bibfnamefont {D.}~\bibnamefont {Voneshen}}, \bibinfo {author}
  {\bibfnamefont {A.~A.}\ \bibnamefont {Tsirlin}}, \bibinfo {author}
  {\bibfnamefont {P.}~\bibnamefont {Gegenwart}}, \ and\ \bibinfo {author}
  {\bibfnamefont {Q.}~\bibnamefont {Zhang}},\ }\bibfield  {title} {\enquote
  {\bibinfo {title} {{Crystalline Electric-Field Randomness in the Triangular
  Lattice Spin-Liquid ${\mathrm{YbMgGaO}}_{4}$}},}\ }\href {\doibase
  10.1103/PhysRevLett.118.107202} {\bibfield  {journal} {\bibinfo  {journal}
  {Phys. Rev. Lett.}\ }\textbf {\bibinfo {volume} {118}},\ \bibinfo {pages}
  {107202} (\bibinfo {year} {2017})}\BibitemShut {NoStop}%
\bibitem [{\citenamefont {Steinhardt}\ \emph {et~al.}(2019)\citenamefont
  {Steinhardt}, \citenamefont {Shi}, \citenamefont {Samarakoon}, \citenamefont
  {Dissanayake}, \citenamefont {Graf}, \citenamefont {Liu}, \citenamefont
  {Zhu}, \citenamefont {Marjerrison}, \citenamefont {Batista},\ and\
  \citenamefont {Haravifard}}]{steinhardt2019field}%
  \BibitemOpen
  \bibfield  {author} {\bibinfo {author} {\bibfnamefont {W.~M.}\ \bibnamefont
  {Steinhardt}}, \bibinfo {author} {\bibfnamefont {Z.}~\bibnamefont {Shi}},
  \bibinfo {author} {\bibfnamefont {A.}~\bibnamefont {Samarakoon}}, \bibinfo
  {author} {\bibfnamefont {S.}~\bibnamefont {Dissanayake}}, \bibinfo {author}
  {\bibfnamefont {D.}~\bibnamefont {Graf}}, \bibinfo {author} {\bibfnamefont
  {Y.}~\bibnamefont {Liu}}, \bibinfo {author} {\bibfnamefont {W.}~\bibnamefont
  {Zhu}}, \bibinfo {author} {\bibfnamefont {C.}~\bibnamefont {Marjerrison}},
  \bibinfo {author} {\bibfnamefont {C.~D.}\ \bibnamefont {Batista}}, \ and\
  \bibinfo {author} {\bibfnamefont {S.}~\bibnamefont {Haravifard}},\ }\bibfield
   {title} {\enquote {\bibinfo {title} {{Field-Induced Phase Transition of the
  Spin Liquid State in Triangular Antiferromagnet YbMgGaO$_4$}},}\ }\href
  {https://arxiv.org/abs/1902.07825} {\bibfield  {journal} {\bibinfo  {journal}
  {arXiv preprint arXiv:1902.07825}\ } (\bibinfo {year} {2019})}\BibitemShut
  {NoStop}%
\bibitem [{\citenamefont {Li}(2019)}]{li2019ybmggao4}%
  \BibitemOpen
  \bibfield  {author} {\bibinfo {author} {\bibfnamefont {Y.}~\bibnamefont
  {Li}},\ }\bibfield  {title} {\enquote {\bibinfo {title} {{YbMgGaO$_4$: A
  Triangular-Lattice Quantum Spin Liquid Candidate}},}\ }\href@noop {}
  {\bibfield  {journal} {\bibinfo  {journal} {Advanced Quantum Technologies}\
  }\textbf {\bibinfo {volume} {2}},\ \bibinfo {pages} {1900089} (\bibinfo
  {year} {2019})}\BibitemShut {NoStop}%
\bibitem [{\citenamefont {Shen}\ \emph {et~al.}(2018)\citenamefont {Shen},
  \citenamefont {Li}, \citenamefont {Walker}, \citenamefont {Steffens},
  \citenamefont {Boehm}, \citenamefont {Zhang}, \citenamefont {Shen},
  \citenamefont {Wo}, \citenamefont {Chen},\ and\ \citenamefont
  {Zhao}}]{shen2018fractionalized}%
  \BibitemOpen
  \bibfield  {author} {\bibinfo {author} {\bibfnamefont {Y.}~\bibnamefont
  {Shen}}, \bibinfo {author} {\bibfnamefont {Y.-D.}\ \bibnamefont {Li}},
  \bibinfo {author} {\bibfnamefont {H.}~\bibnamefont {Walker}}, \bibinfo
  {author} {\bibfnamefont {P.}~\bibnamefont {Steffens}}, \bibinfo {author}
  {\bibfnamefont {M.}~\bibnamefont {Boehm}}, \bibinfo {author} {\bibfnamefont
  {X.}~\bibnamefont {Zhang}}, \bibinfo {author} {\bibfnamefont
  {S.}~\bibnamefont {Shen}}, \bibinfo {author} {\bibfnamefont {H.}~\bibnamefont
  {Wo}}, \bibinfo {author} {\bibfnamefont {G.}~\bibnamefont {Chen}}, \ and\
  \bibinfo {author} {\bibfnamefont {J.}~\bibnamefont {Zhao}},\ }\bibfield
  {title} {\enquote {\bibinfo {title} {{Fractionalized excitations in the
  partially magnetized spin liquid candidate YbMgGaO$_4$}},}\ }\href {\doibase
  https://doi.org/10.1038/s41467-018-06588-1} {\bibfield  {journal} {\bibinfo
  {journal} {Nat. Commun.}\ }\textbf {\bibinfo {volume} {9}},\ \bibinfo {pages}
  {4138} (\bibinfo {year} {2018})}\BibitemShut {NoStop}%
\bibitem [{\citenamefont {Zhu}\ \emph {et~al.}(2017)\citenamefont {Zhu},
  \citenamefont {Maksimov}, \citenamefont {White},\ and\ \citenamefont
  {Chernyshev}}]{zhu2017disorder}%
  \BibitemOpen
  \bibfield  {author} {\bibinfo {author} {\bibfnamefont {Z.}~\bibnamefont
  {Zhu}}, \bibinfo {author} {\bibfnamefont {P.~A.}\ \bibnamefont {Maksimov}},
  \bibinfo {author} {\bibfnamefont {S.~R.}\ \bibnamefont {White}}, \ and\
  \bibinfo {author} {\bibfnamefont {A.~L.}\ \bibnamefont {Chernyshev}},\
  }\bibfield  {title} {\enquote {\bibinfo {title} {{Disorder-Induced Mimicry of
  a Spin Liquid in ${\mathrm{YbMgGaO}}_{4}$}},}\ }\href {\doibase
  10.1103/PhysRevLett.119.157201} {\bibfield  {journal} {\bibinfo  {journal}
  {Phys. Rev. Lett.}\ }\textbf {\bibinfo {volume} {119}},\ \bibinfo {pages}
  {157201} (\bibinfo {year} {2017})}\BibitemShut {NoStop}%
\bibitem [{\citenamefont {Zhu}\ \emph {et~al.}(2018)\citenamefont {Zhu},
  \citenamefont {Maksimov}, \citenamefont {White},\ and\ \citenamefont
  {Chernyshev}}]{zhu2018topography}%
  \BibitemOpen
  \bibfield  {author} {\bibinfo {author} {\bibfnamefont {Z.}~\bibnamefont
  {Zhu}}, \bibinfo {author} {\bibfnamefont {P.~A.}\ \bibnamefont {Maksimov}},
  \bibinfo {author} {\bibfnamefont {S.~R.}\ \bibnamefont {White}}, \ and\
  \bibinfo {author} {\bibfnamefont {A.~L.}\ \bibnamefont {Chernyshev}},\
  }\bibfield  {title} {\enquote {\bibinfo {title} {{Topography of Spin Liquids
  on a Triangular Lattice}},}\ }\href {\doibase 10.1103/PhysRevLett.120.207203}
  {\bibfield  {journal} {\bibinfo  {journal} {Phys. Rev. Lett.}\ }\textbf
  {\bibinfo {volume} {120}},\ \bibinfo {pages} {207203} (\bibinfo {year}
  {2018})}\BibitemShut {NoStop}%
\bibitem [{\citenamefont {Kimchi}, \citenamefont {Nahum},\ and\ \citenamefont
  {Senthil}(2018)}]{kimchi2018valence}%
  \BibitemOpen
  \bibfield  {author} {\bibinfo {author} {\bibfnamefont {I.}~\bibnamefont
  {Kimchi}}, \bibinfo {author} {\bibfnamefont {A.}~\bibnamefont {Nahum}}, \
  and\ \bibinfo {author} {\bibfnamefont {T.}~\bibnamefont {Senthil}},\
  }\bibfield  {title} {\enquote {\bibinfo {title} {{Valence Bonds in Random
  Quantum Magnets: Theory and Application to ${\mathrm{YbMgGaO}}_{4}$}},}\
  }\href {\doibase 10.1103/PhysRevX.8.031028} {\bibfield  {journal} {\bibinfo
  {journal} {Phys. Rev. X}\ }\textbf {\bibinfo {volume} {8}},\ \bibinfo {pages}
  {031028} (\bibinfo {year} {2018})}\BibitemShut {NoStop}%
\bibitem [{\citenamefont {Liu}\ \emph {et~al.}(2018)\citenamefont {Liu},
  \citenamefont {Zhang}, \citenamefont {Ji}, \citenamefont {Liu}, \citenamefont
  {Li}, \citenamefont {Wang}, \citenamefont {Lei}, \citenamefont {Chen},\ and\
  \citenamefont {Zhang}}]{liu2018rare}%
  \BibitemOpen
  \bibfield  {author} {\bibinfo {author} {\bibfnamefont {W.}~\bibnamefont
  {Liu}}, \bibinfo {author} {\bibfnamefont {Z.}~\bibnamefont {Zhang}}, \bibinfo
  {author} {\bibfnamefont {J.}~\bibnamefont {Ji}}, \bibinfo {author}
  {\bibfnamefont {Y.}~\bibnamefont {Liu}}, \bibinfo {author} {\bibfnamefont
  {J.}~\bibnamefont {Li}}, \bibinfo {author} {\bibfnamefont {X.}~\bibnamefont
  {Wang}}, \bibinfo {author} {\bibfnamefont {H.}~\bibnamefont {Lei}}, \bibinfo
  {author} {\bibfnamefont {G.}~\bibnamefont {Chen}}, \ and\ \bibinfo {author}
  {\bibfnamefont {Q.}~\bibnamefont {Zhang}},\ }\bibfield  {title} {\enquote
  {\bibinfo {title} {{Rare-Earth Chalcogenides: A Large Family of Triangular
  Lattice Spin Liquid Candidates}},}\ }\href {\doibase
  10.1088/0256-307x/35/11/117501} {\bibfield  {journal} {\bibinfo  {journal}
  {Chinese Phys. Lett.}\ }\textbf {\bibinfo {volume} {35}},\ \bibinfo {pages}
  {117501} (\bibinfo {year} {2018})}\BibitemShut {NoStop}%
\bibitem [{\citenamefont {Xing}\ \emph
  {et~al.}(2019{\natexlab{a}})\citenamefont {Xing}, \citenamefont {Sanjeewa},
  \citenamefont {Kim}, \citenamefont {Stewart}, \citenamefont {Du},
  \citenamefont {Reboredo}, \citenamefont {Custelcean},\ and\ \citenamefont
  {Sefat}}]{xing2019}%
  \BibitemOpen
  \bibfield  {author} {\bibinfo {author} {\bibfnamefont {J.}~\bibnamefont
  {Xing}}, \bibinfo {author} {\bibfnamefont {L.~D.}\ \bibnamefont {Sanjeewa}},
  \bibinfo {author} {\bibfnamefont {J.}~\bibnamefont {Kim}}, \bibinfo {author}
  {\bibfnamefont {G.}~\bibnamefont {Stewart}}, \bibinfo {author} {\bibfnamefont
  {M.-H.}\ \bibnamefont {Du}}, \bibinfo {author} {\bibfnamefont {F.~A.}\
  \bibnamefont {Reboredo}}, \bibinfo {author} {\bibfnamefont {R.}~\bibnamefont
  {Custelcean}}, \ and\ \bibinfo {author} {\bibfnamefont {A.~S.}\ \bibnamefont
  {Sefat}},\ }\href {\doibase 10.1021/acsmaterialslett.9b00464} {\bibfield
  {journal} {\bibinfo  {journal} {ACS Materials Lett.}\ ,\ \bibinfo {pages}
  {71--75}} (\bibinfo {year} {2019}{\natexlab{a}})}\BibitemShut {NoStop}%
\bibitem [{\citenamefont {Xing}\ \emph
  {et~al.}(2019{\natexlab{b}})\citenamefont {Xing}, \citenamefont {Sanjeewa},
  \citenamefont {Kim}, \citenamefont {Meier}, \citenamefont {May},
  \citenamefont {Zheng}, \citenamefont {Custelcean}, \citenamefont {Stewart},\
  and\ \citenamefont {Sefat}}]{jie2019naerse}%
  \BibitemOpen
  \bibfield  {author} {\bibinfo {author} {\bibfnamefont {J.}~\bibnamefont
  {Xing}}, \bibinfo {author} {\bibfnamefont {L.~D.}\ \bibnamefont {Sanjeewa}},
  \bibinfo {author} {\bibfnamefont {J.}~\bibnamefont {Kim}}, \bibinfo {author}
  {\bibfnamefont {W.~R.}\ \bibnamefont {Meier}}, \bibinfo {author}
  {\bibfnamefont {A.~F.}\ \bibnamefont {May}}, \bibinfo {author} {\bibfnamefont
  {Q.}~\bibnamefont {Zheng}}, \bibinfo {author} {\bibfnamefont
  {R.}~\bibnamefont {Custelcean}}, \bibinfo {author} {\bibfnamefont {G.~R.}\
  \bibnamefont {Stewart}}, \ and\ \bibinfo {author} {\bibfnamefont {A.~S.}\
  \bibnamefont {Sefat}},\ }\bibfield  {title} {\enquote {\bibinfo {title}
  {{Synthesis, magnetization, and heat capacity of triangular lattice materials
  ${\mathrm{NaErSe}}_{2}$ and ${\mathrm{KErSe}}_{2}$}},}\ }\href {\doibase
  10.1103/PhysRevMaterials.3.114413} {\bibfield  {journal} {\bibinfo  {journal}
  {Phys. Rev. Materials}\ }\textbf {\bibinfo {volume} {3}},\ \bibinfo {pages}
  {114413} (\bibinfo {year} {2019}{\natexlab{b}})}\BibitemShut {NoStop}%
\bibitem [{\citenamefont {Zangeneh}\ \emph {et~al.}(2019)\citenamefont
  {Zangeneh}, \citenamefont {Avdoshenko}, \citenamefont {van~den Brink},\ and\
  \citenamefont {Hozoi}}]{zangeneh2019single}%
  \BibitemOpen
  \bibfield  {author} {\bibinfo {author} {\bibfnamefont {Z.}~\bibnamefont
  {Zangeneh}}, \bibinfo {author} {\bibfnamefont {S.}~\bibnamefont
  {Avdoshenko}}, \bibinfo {author} {\bibfnamefont {J.}~\bibnamefont {van~den
  Brink}}, \ and\ \bibinfo {author} {\bibfnamefont {L.}~\bibnamefont {Hozoi}},\
  }\bibfield  {title} {\enquote {\bibinfo {title} {{Single-site magnetic
  anisotropy governed by interlayer cation charge imbalance in
  triangular-lattice $A\mathrm{Yb}{X}_{2}$}},}\ }\href@noop {} {\bibfield
  {journal} {\bibinfo  {journal} {Phys. Rev. B}\ }\textbf {\bibinfo {volume}
  {100}},\ \bibinfo {pages} {174436} (\bibinfo {year} {2019})}\BibitemShut
  {NoStop}%
\bibitem [{\citenamefont {Scheie}\ \emph {et~al.}(2020)\citenamefont {Scheie},
  \citenamefont {Garlea}, \citenamefont {Sanjeewa}, \citenamefont {Xing},\ and\
  \citenamefont {Sefat}}]{scheie2020crystal}%
  \BibitemOpen
  \bibfield  {author} {\bibinfo {author} {\bibfnamefont {A.}~\bibnamefont
  {Scheie}}, \bibinfo {author} {\bibfnamefont {V.~O.}\ \bibnamefont {Garlea}},
  \bibinfo {author} {\bibfnamefont {L.~D.}\ \bibnamefont {Sanjeewa}}, \bibinfo
  {author} {\bibfnamefont {J.}~\bibnamefont {Xing}}, \ and\ \bibinfo {author}
  {\bibfnamefont {A.~S.}\ \bibnamefont {Sefat}},\ }\bibfield  {title} {\enquote
  {\bibinfo {title} {{Crystal-field Hamiltonian and anisotropy in
  ${\mathrm{KErSe}}_{2}$ and ${\mathrm{CsErSe}}_{2}$}},}\ }\href@noop {}
  {\bibfield  {journal} {\bibinfo  {journal} {Phys. Rev. B}\ }\textbf {\bibinfo
  {volume} {101}},\ \bibinfo {pages} {144432} (\bibinfo {year}
  {2020})}\BibitemShut {NoStop}%
\bibitem [{\citenamefont {Bordelon}\ \emph {et~al.}(2021)\citenamefont
  {Bordelon}, \citenamefont {Liu}, \citenamefont {Posthuma}, \citenamefont
  {Kenney}, \citenamefont {Graf}, \citenamefont {Butch}, \citenamefont
  {Banerjee}, \citenamefont {Calder}, \citenamefont {Balents},\ and\
  \citenamefont {Wilson}}]{bordelon2021frustrated}%
  \BibitemOpen
  \bibfield  {author} {\bibinfo {author} {\bibfnamefont {M.~M.}\ \bibnamefont
  {Bordelon}}, \bibinfo {author} {\bibfnamefont {C.}~\bibnamefont {Liu}},
  \bibinfo {author} {\bibfnamefont {L.}~\bibnamefont {Posthuma}}, \bibinfo
  {author} {\bibfnamefont {E.}~\bibnamefont {Kenney}}, \bibinfo {author}
  {\bibfnamefont {M.}~\bibnamefont {Graf}}, \bibinfo {author} {\bibfnamefont
  {N.~P.}\ \bibnamefont {Butch}}, \bibinfo {author} {\bibfnamefont
  {A.}~\bibnamefont {Banerjee}}, \bibinfo {author} {\bibfnamefont
  {S.}~\bibnamefont {Calder}}, \bibinfo {author} {\bibfnamefont
  {L.}~\bibnamefont {Balents}}, \ and\ \bibinfo {author} {\bibfnamefont
  {S.~D.}\ \bibnamefont {Wilson}},\ }\bibfield  {title} {\enquote {\bibinfo
  {title} {{Frustrated Heisenberg J$_1$-J$_2$ model within the stretched
  diamond lattice of LiYbO$_2$}},}\ }\href@noop {} {\bibfield  {journal}
  {\bibinfo  {journal} {Physical Review B}\ }\textbf {\bibinfo {volume}
  {103}},\ \bibinfo {pages} {014420} (\bibinfo {year} {2021})}\BibitemShut
  {NoStop}%
\bibitem [{\citenamefont {Gao}\ \emph {et~al.}(2020)\citenamefont {Gao},
  \citenamefont {Xiao}, \citenamefont {Kamazawa}, \citenamefont {Ikeuchi},
  \citenamefont {Biner}, \citenamefont {Kr{\"a}mer}, \citenamefont
  {R{\"u}egg},\ and\ \citenamefont {Arima}}]{gao2020crystal}%
  \BibitemOpen
  \bibfield  {author} {\bibinfo {author} {\bibfnamefont {S.}~\bibnamefont
  {Gao}}, \bibinfo {author} {\bibfnamefont {F.}~\bibnamefont {Xiao}}, \bibinfo
  {author} {\bibfnamefont {K.}~\bibnamefont {Kamazawa}}, \bibinfo {author}
  {\bibfnamefont {K.}~\bibnamefont {Ikeuchi}}, \bibinfo {author} {\bibfnamefont
  {D.}~\bibnamefont {Biner}}, \bibinfo {author} {\bibfnamefont {K.~W.}\
  \bibnamefont {Kr{\"a}mer}}, \bibinfo {author} {\bibfnamefont
  {C.}~\bibnamefont {R{\"u}egg}}, \ and\ \bibinfo {author} {\bibfnamefont
  {T.-h.}\ \bibnamefont {Arima}},\ }\bibfield  {title} {\enquote {\bibinfo
  {title} {{Crystal electric field excitations in the quantum spin liquid
  candidate NaErS$_2$}},}\ }\href@noop {} {\bibfield  {journal} {\bibinfo
  {journal} {Physical Review B}\ }\textbf {\bibinfo {volume} {102}},\ \bibinfo
  {pages} {024424} (\bibinfo {year} {2020})}\BibitemShut {NoStop}%
\bibitem [{\citenamefont {Xing}\ \emph {et~al.}(2021)\citenamefont {Xing},
  \citenamefont {Taddei}, \citenamefont {Sanjeewa}, \citenamefont {Fishman},
  \citenamefont {Daum}, \citenamefont {Mourigal}, \citenamefont {dela Cruz},\
  and\ \citenamefont {Sefat}}]{xing2021stripe}%
  \BibitemOpen
  \bibfield  {author} {\bibinfo {author} {\bibfnamefont {J.}~\bibnamefont
  {Xing}}, \bibinfo {author} {\bibfnamefont {K.~M.}\ \bibnamefont {Taddei}},
  \bibinfo {author} {\bibfnamefont {L.~D.}\ \bibnamefont {Sanjeewa}}, \bibinfo
  {author} {\bibfnamefont {R.~S.}\ \bibnamefont {Fishman}}, \bibinfo {author}
  {\bibfnamefont {M.}~\bibnamefont {Daum}}, \bibinfo {author} {\bibfnamefont
  {M.}~\bibnamefont {Mourigal}}, \bibinfo {author} {\bibfnamefont
  {C.}~\bibnamefont {dela Cruz}}, \ and\ \bibinfo {author} {\bibfnamefont
  {A.~S.}\ \bibnamefont {Sefat}},\ }\bibfield  {title} {\enquote {\bibinfo
  {title} {{Stripe antiferromagnetic ground state of the ideal triangular
  lattice compound KErSe$_2$}},}\ }\href@noop {} {\bibfield  {journal}
  {\bibinfo  {journal} {Physical Review B}\ }\textbf {\bibinfo {volume}
  {103}},\ \bibinfo {pages} {144413} (\bibinfo {year} {2021})}\BibitemShut
  {NoStop}%
\bibitem [{\citenamefont {Baenitz}\ \emph {et~al.}(2018)\citenamefont
  {Baenitz}, \citenamefont {Schlender}, \citenamefont {Sichelschmidt},
  \citenamefont {Onykiienko}, \citenamefont {Zangeneh}, \citenamefont
  {Ranjith}, \citenamefont {Sarkar}, \citenamefont {Hozoi}, \citenamefont
  {Walker}, \citenamefont {Orain}, \citenamefont {Yasuoka}, \citenamefont
  {van~den Brink}, \citenamefont {Klauss}, \citenamefont {Inosov},\ and\
  \citenamefont {Doert}}]{baenitz2018naybs}%
  \BibitemOpen
  \bibfield  {author} {\bibinfo {author} {\bibfnamefont {M.}~\bibnamefont
  {Baenitz}}, \bibinfo {author} {\bibfnamefont {P.}~\bibnamefont {Schlender}},
  \bibinfo {author} {\bibfnamefont {J.}~\bibnamefont {Sichelschmidt}}, \bibinfo
  {author} {\bibfnamefont {Y.~A.}\ \bibnamefont {Onykiienko}}, \bibinfo
  {author} {\bibfnamefont {Z.}~\bibnamefont {Zangeneh}}, \bibinfo {author}
  {\bibfnamefont {K.~M.}\ \bibnamefont {Ranjith}}, \bibinfo {author}
  {\bibfnamefont {R.}~\bibnamefont {Sarkar}}, \bibinfo {author} {\bibfnamefont
  {L.}~\bibnamefont {Hozoi}}, \bibinfo {author} {\bibfnamefont {H.~C.}\
  \bibnamefont {Walker}}, \bibinfo {author} {\bibfnamefont {J.-C.}\
  \bibnamefont {Orain}}, \bibinfo {author} {\bibfnamefont {H.}~\bibnamefont
  {Yasuoka}}, \bibinfo {author} {\bibfnamefont {J.}~\bibnamefont {van~den
  Brink}}, \bibinfo {author} {\bibfnamefont {H.~H.}\ \bibnamefont {Klauss}},
  \bibinfo {author} {\bibfnamefont {D.~S.}\ \bibnamefont {Inosov}}, \ and\
  \bibinfo {author} {\bibfnamefont {T.}~\bibnamefont {Doert}},\ }\bibfield
  {title} {\enquote {\bibinfo {title} {{NaYbS$_{2}$: A planar spin-$1/2$
  triangular-lattice magnet and putative spin liquid}},}\ }\href {\doibase
  10.1103/PhysRevB.98.220409} {\bibfield  {journal} {\bibinfo  {journal} {Phys.
  Rev. B}\ }\textbf {\bibinfo {volume} {98}},\ \bibinfo {pages} {220409}
  (\bibinfo {year} {2018})}\BibitemShut {NoStop}%
\bibitem [{\citenamefont {Bordelon}\ \emph {et~al.}(2020)\citenamefont
  {Bordelon}, \citenamefont {Liu}, \citenamefont {Posthuma}, \citenamefont
  {Sarte}, \citenamefont {Butch}, \citenamefont {Pajerowski}, \citenamefont
  {Banerjee}, \citenamefont {Balents},\ and\ \citenamefont
  {Wilson}}]{bordelon2020spin}%
  \BibitemOpen
  \bibfield  {author} {\bibinfo {author} {\bibfnamefont {M.~M.}\ \bibnamefont
  {Bordelon}}, \bibinfo {author} {\bibfnamefont {C.}~\bibnamefont {Liu}},
  \bibinfo {author} {\bibfnamefont {L.}~\bibnamefont {Posthuma}}, \bibinfo
  {author} {\bibfnamefont {P.}~\bibnamefont {Sarte}}, \bibinfo {author}
  {\bibfnamefont {N.}~\bibnamefont {Butch}}, \bibinfo {author} {\bibfnamefont
  {D.~M.}\ \bibnamefont {Pajerowski}}, \bibinfo {author} {\bibfnamefont
  {A.}~\bibnamefont {Banerjee}}, \bibinfo {author} {\bibfnamefont
  {L.}~\bibnamefont {Balents}}, \ and\ \bibinfo {author} {\bibfnamefont
  {S.~D.}\ \bibnamefont {Wilson}},\ }\bibfield  {title} {\enquote {\bibinfo
  {title} {{Spin excitations in the frustrated triangular lattice
  antiferromagnet NaYbO$_2$}},}\ }\href@noop {} {\bibfield  {journal} {\bibinfo
   {journal} {Physical Review B}\ }\textbf {\bibinfo {volume} {101}},\ \bibinfo
  {pages} {224427} (\bibinfo {year} {2020})}\BibitemShut {NoStop}%
\bibitem [{\citenamefont {Zhang}\ \emph
  {et~al.}(2021{\natexlab{a}})\citenamefont {Zhang}, \citenamefont {Ma},
  \citenamefont {Li}, \citenamefont {Wang}, \citenamefont {Adroja},
  \citenamefont {Perring}, \citenamefont {Liu}, \citenamefont {Jin},
  \citenamefont {Ji}, \citenamefont {Wang} \emph
  {et~al.}}]{zhang2021crystalline}%
  \BibitemOpen
  \bibfield  {author} {\bibinfo {author} {\bibfnamefont {Z.}~\bibnamefont
  {Zhang}}, \bibinfo {author} {\bibfnamefont {X.}~\bibnamefont {Ma}}, \bibinfo
  {author} {\bibfnamefont {J.}~\bibnamefont {Li}}, \bibinfo {author}
  {\bibfnamefont {G.}~\bibnamefont {Wang}}, \bibinfo {author} {\bibfnamefont
  {D.}~\bibnamefont {Adroja}}, \bibinfo {author} {\bibfnamefont
  {T.}~\bibnamefont {Perring}}, \bibinfo {author} {\bibfnamefont
  {W.}~\bibnamefont {Liu}}, \bibinfo {author} {\bibfnamefont {F.}~\bibnamefont
  {Jin}}, \bibinfo {author} {\bibfnamefont {J.}~\bibnamefont {Ji}}, \bibinfo
  {author} {\bibfnamefont {Y.}~\bibnamefont {Wang}},  \emph {et~al.},\
  }\bibfield  {title} {\enquote {\bibinfo {title} {{Crystalline electric field
  excitations in the quantum spin liquid candidate NaYbSe$_2$}},}\ }\href@noop
  {} {\bibfield  {journal} {\bibinfo  {journal} {Physical Review B}\ }\textbf
  {\bibinfo {volume} {103}},\ \bibinfo {pages} {035144} (\bibinfo {year}
  {2021}{\natexlab{a}})}\BibitemShut {NoStop}%
\bibitem [{\citenamefont {Sichelschmidt}\ \emph {et~al.}(2020)\citenamefont
  {Sichelschmidt}, \citenamefont {Schmidt}, \citenamefont {Schlender},
  \citenamefont {Khim}, \citenamefont {Doert},\ and\ \citenamefont
  {Baenitz}}]{sichelschmidt2020effective}%
  \BibitemOpen
  \bibfield  {author} {\bibinfo {author} {\bibfnamefont {J.}~\bibnamefont
  {Sichelschmidt}}, \bibinfo {author} {\bibfnamefont {B.}~\bibnamefont
  {Schmidt}}, \bibinfo {author} {\bibfnamefont {P.}~\bibnamefont {Schlender}},
  \bibinfo {author} {\bibfnamefont {S.}~\bibnamefont {Khim}}, \bibinfo {author}
  {\bibfnamefont {T.}~\bibnamefont {Doert}}, \ and\ \bibinfo {author}
  {\bibfnamefont {M.}~\bibnamefont {Baenitz}},\ }\bibfield  {title} {\enquote
  {\bibinfo {title} {{Effective Spin-1/2 Moments on a Yb$^{3+}$ Triangular
  Lattice: An ESR Study}},}\ }in\ \href@noop {} {\emph {\bibinfo {booktitle}
  {Proceedings of the International Conference on Strongly Correlated Electron
  Systems (SCES2019)}}}\ (\bibinfo {year} {2020})\ p.\ \bibinfo {pages}
  {011096}\BibitemShut {NoStop}%
\bibitem [{\citenamefont {Guo}\ \emph {et~al.}(2020)\citenamefont {Guo},
  \citenamefont {Zhao}, \citenamefont {Ohira-Kawamura}, \citenamefont {Ling},
  \citenamefont {Wang}, \citenamefont {He}, \citenamefont {Nakajima},
  \citenamefont {Li},\ and\ \citenamefont {Zhang}}]{guo2020magnetic}%
  \BibitemOpen
  \bibfield  {author} {\bibinfo {author} {\bibfnamefont {J.}~\bibnamefont
  {Guo}}, \bibinfo {author} {\bibfnamefont {X.}~\bibnamefont {Zhao}}, \bibinfo
  {author} {\bibfnamefont {S.}~\bibnamefont {Ohira-Kawamura}}, \bibinfo
  {author} {\bibfnamefont {L.}~\bibnamefont {Ling}}, \bibinfo {author}
  {\bibfnamefont {J.}~\bibnamefont {Wang}}, \bibinfo {author} {\bibfnamefont
  {L.}~\bibnamefont {He}}, \bibinfo {author} {\bibfnamefont {K.}~\bibnamefont
  {Nakajima}}, \bibinfo {author} {\bibfnamefont {B.}~\bibnamefont {Li}}, \ and\
  \bibinfo {author} {\bibfnamefont {Z.}~\bibnamefont {Zhang}},\ }\bibfield
  {title} {\enquote {\bibinfo {title} {{Magnetic-field and composition tuned
  antiferromagnetic instability in the quantum spin-liquid candidate
  NaYbO$_2$}},}\ }\href@noop {} {\bibfield  {journal} {\bibinfo  {journal}
  {Physical Review Materials}\ }\textbf {\bibinfo {volume} {4}},\ \bibinfo
  {pages} {064410} (\bibinfo {year} {2020})}\BibitemShut {NoStop}%
\bibitem [{\citenamefont {Zhang}\ \emph
  {et~al.}(2021{\natexlab{b}})\citenamefont {Zhang}, \citenamefont {Li},
  \citenamefont {Liu}, \citenamefont {Zhang}, \citenamefont {Ji}, \citenamefont
  {Jin}, \citenamefont {Chen}, \citenamefont {Wang}, \citenamefont {Wang},
  \citenamefont {Ma} \emph {et~al.}}]{zhang2021effective}%
  \BibitemOpen
  \bibfield  {author} {\bibinfo {author} {\bibfnamefont {Z.}~\bibnamefont
  {Zhang}}, \bibinfo {author} {\bibfnamefont {J.}~\bibnamefont {Li}}, \bibinfo
  {author} {\bibfnamefont {W.}~\bibnamefont {Liu}}, \bibinfo {author}
  {\bibfnamefont {Z.}~\bibnamefont {Zhang}}, \bibinfo {author} {\bibfnamefont
  {J.}~\bibnamefont {Ji}}, \bibinfo {author} {\bibfnamefont {F.}~\bibnamefont
  {Jin}}, \bibinfo {author} {\bibfnamefont {R.}~\bibnamefont {Chen}}, \bibinfo
  {author} {\bibfnamefont {J.}~\bibnamefont {Wang}}, \bibinfo {author}
  {\bibfnamefont {X.}~\bibnamefont {Wang}}, \bibinfo {author} {\bibfnamefont
  {J.}~\bibnamefont {Ma}},  \emph {et~al.},\ }\bibfield  {title} {\enquote
  {\bibinfo {title} {{Effective magnetic Hamiltonian at finite temperatures for
  rare-earth chalcogenides}},}\ }\href@noop {} {\bibfield  {journal} {\bibinfo
  {journal} {Physical Review B}\ }\textbf {\bibinfo {volume} {103}},\ \bibinfo
  {pages} {184419} (\bibinfo {year} {2021}{\natexlab{b}})}\BibitemShut
  {NoStop}%
\bibitem [{\citenamefont {Schmidt}\ \emph {et~al.}(2021)\citenamefont
  {Schmidt}, \citenamefont {Sichelschmidt}, \citenamefont {Ranjith},
  \citenamefont {Doert},\ and\ \citenamefont {Baenitz}}]{schmidt2021Yb}%
  \BibitemOpen
  \bibfield  {author} {\bibinfo {author} {\bibfnamefont {B.}~\bibnamefont
  {Schmidt}}, \bibinfo {author} {\bibfnamefont {J.}~\bibnamefont
  {Sichelschmidt}}, \bibinfo {author} {\bibfnamefont {K.~M.}\ \bibnamefont
  {Ranjith}}, \bibinfo {author} {\bibfnamefont {T.}~\bibnamefont {Doert}}, \
  and\ \bibinfo {author} {\bibfnamefont {M.}~\bibnamefont {Baenitz}},\
  }\bibfield  {title} {\enquote {\bibinfo {title} {{Yb delafossites: Unique
  exchange frustration of $4f$ spin-$\frac{1}{2}$ moments on a perfect
  triangular lattice}},}\ }\href {\doibase 10.1103/PhysRevB.103.214445}
  {\bibfield  {journal} {\bibinfo  {journal} {Phys. Rev. B}\ }\textbf {\bibinfo
  {volume} {103}},\ \bibinfo {pages} {214445} (\bibinfo {year}
  {2021})}\BibitemShut {NoStop}%
\bibitem [{\citenamefont {Ranjith}\ \emph
  {et~al.}(2019{\natexlab{a}})\citenamefont {Ranjith}, \citenamefont
  {Dmytriieva}, \citenamefont {Khim}, \citenamefont {Sichelschmidt},
  \citenamefont {Luther}, \citenamefont {Ehlers}, \citenamefont {Yasuoka},
  \citenamefont {Wosnitza}, \citenamefont {Tsirlin}, \citenamefont {K\"uhne},\
  and\ \citenamefont {Baenitz}}]{ranjith2019field}%
  \BibitemOpen
  \bibfield  {author} {\bibinfo {author} {\bibfnamefont {K.~M.}\ \bibnamefont
  {Ranjith}}, \bibinfo {author} {\bibfnamefont {D.}~\bibnamefont {Dmytriieva}},
  \bibinfo {author} {\bibfnamefont {S.}~\bibnamefont {Khim}}, \bibinfo {author}
  {\bibfnamefont {J.}~\bibnamefont {Sichelschmidt}}, \bibinfo {author}
  {\bibfnamefont {S.}~\bibnamefont {Luther}}, \bibinfo {author} {\bibfnamefont
  {D.}~\bibnamefont {Ehlers}}, \bibinfo {author} {\bibfnamefont
  {H.}~\bibnamefont {Yasuoka}}, \bibinfo {author} {\bibfnamefont
  {J.}~\bibnamefont {Wosnitza}}, \bibinfo {author} {\bibfnamefont {A.~A.}\
  \bibnamefont {Tsirlin}}, \bibinfo {author} {\bibfnamefont {H.}~\bibnamefont
  {K\"uhne}}, \ and\ \bibinfo {author} {\bibfnamefont {M.}~\bibnamefont
  {Baenitz}},\ }\bibfield  {title} {\enquote {\bibinfo {title} {{Field-induced
  instability of the quantum spin liquid ground state in the
  ${J}_{\mathrm{eff}}=\frac{1}{2}$ triangular-lattice compound
  ${\mathrm{NaYbO}}_{2}$}},}\ }\href {\doibase 10.1103/PhysRevB.99.180401}
  {\bibfield  {journal} {\bibinfo  {journal} {Phys. Rev. B}\ }\textbf {\bibinfo
  {volume} {99}},\ \bibinfo {pages} {180401} (\bibinfo {year}
  {2019}{\natexlab{a}})}\BibitemShut {NoStop}%
\bibitem [{\citenamefont {Ranjith}\ \emph
  {et~al.}(2019{\natexlab{b}})\citenamefont {Ranjith}, \citenamefont {Luther},
  \citenamefont {Reimann}, \citenamefont {Schmidt}, \citenamefont {Schlender},
  \citenamefont {Sichelschmidt}, \citenamefont {Yasuoka}, \citenamefont
  {Strydom}, \citenamefont {Skourski}, \citenamefont {Wosnitza}, \citenamefont
  {K\"uhne}, \citenamefont {Doert},\ and\ \citenamefont
  {Baenitz}}]{Ranjith2019naybse}%
  \BibitemOpen
  \bibfield  {author} {\bibinfo {author} {\bibfnamefont {K.~M.}\ \bibnamefont
  {Ranjith}}, \bibinfo {author} {\bibfnamefont {S.}~\bibnamefont {Luther}},
  \bibinfo {author} {\bibfnamefont {T.}~\bibnamefont {Reimann}}, \bibinfo
  {author} {\bibfnamefont {B.}~\bibnamefont {Schmidt}}, \bibinfo {author}
  {\bibfnamefont {P.}~\bibnamefont {Schlender}}, \bibinfo {author}
  {\bibfnamefont {J.}~\bibnamefont {Sichelschmidt}}, \bibinfo {author}
  {\bibfnamefont {H.}~\bibnamefont {Yasuoka}}, \bibinfo {author} {\bibfnamefont
  {A.~M.}\ \bibnamefont {Strydom}}, \bibinfo {author} {\bibfnamefont
  {Y.}~\bibnamefont {Skourski}}, \bibinfo {author} {\bibfnamefont
  {J.}~\bibnamefont {Wosnitza}}, \bibinfo {author} {\bibfnamefont
  {H.}~\bibnamefont {K\"uhne}}, \bibinfo {author} {\bibfnamefont
  {T.}~\bibnamefont {Doert}}, \ and\ \bibinfo {author} {\bibfnamefont
  {M.}~\bibnamefont {Baenitz}},\ }\bibfield  {title} {\enquote {\bibinfo
  {title} {Anisotropic field-induced ordering in the triangular-lattice quantum
  spin liquid ${\mathrm{naybse}}_{2}$},}\ }\href@noop {} {\bibfield  {journal}
  {\bibinfo  {journal} {Phys. Rev. B}\ }\textbf {\bibinfo {volume} {100}},\
  \bibinfo {pages} {224417} (\bibinfo {year} {2019}{\natexlab{b}})}\BibitemShut
  {NoStop}%
\bibitem [{\citenamefont {Xing}\ \emph
  {et~al.}(2019{\natexlab{c}})\citenamefont {Xing}, \citenamefont {Sanjeewa},
  \citenamefont {Kim}, \citenamefont {Stewart}, \citenamefont {Podlesnyak},\
  and\ \citenamefont {Sefat}}]{PhysRevB.100.220407}%
  \BibitemOpen
  \bibfield  {author} {\bibinfo {author} {\bibfnamefont {J.}~\bibnamefont
  {Xing}}, \bibinfo {author} {\bibfnamefont {L.~D.}\ \bibnamefont {Sanjeewa}},
  \bibinfo {author} {\bibfnamefont {J.}~\bibnamefont {Kim}}, \bibinfo {author}
  {\bibfnamefont {G.~R.}\ \bibnamefont {Stewart}}, \bibinfo {author}
  {\bibfnamefont {A.}~\bibnamefont {Podlesnyak}}, \ and\ \bibinfo {author}
  {\bibfnamefont {A.~S.}\ \bibnamefont {Sefat}},\ }\bibfield  {title} {\enquote
  {\bibinfo {title} {{Field-induced magnetic transition and spin fluctuations
  in the quantum spin-liquid candidate ${\mathrm{CsYbSe}}_{2}$}},}\ }\href
  {\doibase 10.1103/PhysRevB.100.220407} {\bibfield  {journal} {\bibinfo
  {journal} {Phys. Rev. B}\ }\textbf {\bibinfo {volume} {100}},\ \bibinfo
  {pages} {220407} (\bibinfo {year} {2019}{\natexlab{c}})}\BibitemShut
  {NoStop}%
\bibitem [{\citenamefont {Ding}\ \emph {et~al.}(2019)\citenamefont {Ding},
  \citenamefont {Manuel}, \citenamefont {Bachus}, \citenamefont {Gru\ss{}ler},
  \citenamefont {Gegenwart}, \citenamefont {Singleton}, \citenamefont
  {Johnson}, \citenamefont {Walker}, \citenamefont {Adroja}, \citenamefont
  {Hillier},\ and\ \citenamefont {Tsirlin}}]{ding2019gapless}%
  \BibitemOpen
  \bibfield  {author} {\bibinfo {author} {\bibfnamefont {L.}~\bibnamefont
  {Ding}}, \bibinfo {author} {\bibfnamefont {P.}~\bibnamefont {Manuel}},
  \bibinfo {author} {\bibfnamefont {S.}~\bibnamefont {Bachus}}, \bibinfo
  {author} {\bibfnamefont {F.}~\bibnamefont {Gru\ss{}ler}}, \bibinfo {author}
  {\bibfnamefont {P.}~\bibnamefont {Gegenwart}}, \bibinfo {author}
  {\bibfnamefont {J.}~\bibnamefont {Singleton}}, \bibinfo {author}
  {\bibfnamefont {R.~D.}\ \bibnamefont {Johnson}}, \bibinfo {author}
  {\bibfnamefont {H.~C.}\ \bibnamefont {Walker}}, \bibinfo {author}
  {\bibfnamefont {D.~T.}\ \bibnamefont {Adroja}}, \bibinfo {author}
  {\bibfnamefont {A.~D.}\ \bibnamefont {Hillier}}, \ and\ \bibinfo {author}
  {\bibfnamefont {A.~A.}\ \bibnamefont {Tsirlin}},\ }\bibfield  {title}
  {\enquote {\bibinfo {title} {Gapless spin-liquid state in the structurally
  disorder-free triangular antiferromagnet ${\mathrm{naybo}}_{2}$},}\
  }\href@noop {} {\bibfield  {journal} {\bibinfo  {journal} {Phys. Rev. B}\
  }\textbf {\bibinfo {volume} {100}},\ \bibinfo {pages} {144432} (\bibinfo
  {year} {2019})}\BibitemShut {NoStop}%
\bibitem [{\citenamefont {Ma}\ \emph {et~al.}(2020)\citenamefont {Ma},
  \citenamefont {Li}, \citenamefont {Gao}, \citenamefont {Liu}, \citenamefont
  {Zhang}, \citenamefont {Wang}, \citenamefont {Chen}, \citenamefont {Embs},
  \citenamefont {Feng}, \citenamefont {Zhu} \emph {et~al.}}]{ma2020spin}%
  \BibitemOpen
  \bibfield  {author} {\bibinfo {author} {\bibfnamefont {J.}~\bibnamefont
  {Ma}}, \bibinfo {author} {\bibfnamefont {J.}~\bibnamefont {Li}}, \bibinfo
  {author} {\bibfnamefont {Y.~H.}\ \bibnamefont {Gao}}, \bibinfo {author}
  {\bibfnamefont {C.}~\bibnamefont {Liu}}, \bibinfo {author} {\bibfnamefont
  {Z.}~\bibnamefont {Zhang}}, \bibinfo {author} {\bibfnamefont
  {Z.}~\bibnamefont {Wang}}, \bibinfo {author} {\bibfnamefont {R.}~\bibnamefont
  {Chen}}, \bibinfo {author} {\bibfnamefont {J.}~\bibnamefont {Embs}}, \bibinfo
  {author} {\bibfnamefont {E.}~\bibnamefont {Feng}}, \bibinfo {author}
  {\bibfnamefont {F.}~\bibnamefont {Zhu}},  \emph {et~al.},\ }\bibfield
  {title} {\enquote {\bibinfo {title} {{Spin-orbit-coupled triangular-lattice
  spin liquid in rare-earth chalcogenides}},}\ }\href@noop {} {\bibfield
  {journal} {\bibinfo  {journal} {arXiv preprint arXiv:2002.09224}\ } (\bibinfo
  {year} {2020})}\BibitemShut {NoStop}%
\bibitem [{\citenamefont {Ferreira}\ \emph {et~al.}(2020)\citenamefont
  {Ferreira}, \citenamefont {Xing}, \citenamefont {Sanjeewa},\ and\
  \citenamefont {Sefat}}]{ferreira2020frustrated}%
  \BibitemOpen
  \bibfield  {author} {\bibinfo {author} {\bibfnamefont {T.}~\bibnamefont
  {Ferreira}}, \bibinfo {author} {\bibfnamefont {J.}~\bibnamefont {Xing}},
  \bibinfo {author} {\bibfnamefont {L.~D.}\ \bibnamefont {Sanjeewa}}, \ and\
  \bibinfo {author} {\bibfnamefont {A.~S.}\ \bibnamefont {Sefat}},\ }\bibfield
  {title} {\enquote {\bibinfo {title} {{Frustrated Magnetism in Triangular
  Lattice TlYbS$_2$ Crystals Grown via Molten Flux}},}\ }\href@noop {}
  {\bibfield  {journal} {\bibinfo  {journal} {Frontiers in Chemistry}\ }\textbf
  {\bibinfo {volume} {8}},\ \bibinfo {pages} {127} (\bibinfo {year}
  {2020})}\BibitemShut {NoStop}%
\bibitem [{\citenamefont {{K. M. Ranjith and Ph. Schlender and Th. Doert and M.
  Baenitz}}()}]{liybs2}%
  \BibitemOpen
  \bibfield  {author} {\bibinfo {author} {\bibnamefont {{K. M. Ranjith and Ph.
  Schlender and Th. Doert and M. Baenitz}}},\ }\bibfield  {title} {\enquote
  {\bibinfo {title} {{Magnetic Properties of an Effective Spin-1/2 Triangualr
  Lattcie Compound LiYbS$_2$}},}\ }in\ \href@noop {} {\emph {\bibinfo
  {booktitle} {{Proceedings of the International Conference on Strongly
  Correlated Electron Systems (SCES2019)}}}}\BibitemShut {NoStop}%
\bibitem [{\citenamefont {Iizuka}\ \emph {et~al.}(2020)\citenamefont {Iizuka},
  \citenamefont {Michimura}, \citenamefont {Numakura}, \citenamefont
  {Uwatoko},\ and\ \citenamefont {Kosaka}}]{iizuka2020single}%
  \BibitemOpen
  \bibfield  {author} {\bibinfo {author} {\bibfnamefont {R.}~\bibnamefont
  {Iizuka}}, \bibinfo {author} {\bibfnamefont {S.}~\bibnamefont {Michimura}},
  \bibinfo {author} {\bibfnamefont {R.}~\bibnamefont {Numakura}}, \bibinfo
  {author} {\bibfnamefont {Y.}~\bibnamefont {Uwatoko}}, \ and\ \bibinfo
  {author} {\bibfnamefont {M.}~\bibnamefont {Kosaka}},\ }\bibfield  {title}
  {\enquote {\bibinfo {title} {{Single Crystal Growth and Physical Properties
  of Ytterbium Sulfide with Triangular Lattice}},}\ }in\ \href@noop {} {\emph
  {\bibinfo {booktitle} {Proceedings of the International Conference on
  Strongly Correlated Electron Systems (SCES2019)}}}\ (\bibinfo {year} {2020})\
  p.\ \bibinfo {pages} {011097}\BibitemShut {NoStop}%
\bibitem [{\citenamefont {Zhang}\ \emph
  {et~al.}(2020{\natexlab{a}})\citenamefont {Zhang}, \citenamefont {Ma},
  \citenamefont {Li}, \citenamefont {Wang}, \citenamefont {Adroja},
  \citenamefont {Perring}, \citenamefont {Liu}, \citenamefont {Jin},
  \citenamefont {Ji}, \citenamefont {Wang} \emph
  {et~al.}}]{zhang2020crystalline}%
  \BibitemOpen
  \bibfield  {author} {\bibinfo {author} {\bibfnamefont {Z.}~\bibnamefont
  {Zhang}}, \bibinfo {author} {\bibfnamefont {X.}~\bibnamefont {Ma}}, \bibinfo
  {author} {\bibfnamefont {J.}~\bibnamefont {Li}}, \bibinfo {author}
  {\bibfnamefont {G.}~\bibnamefont {Wang}}, \bibinfo {author} {\bibfnamefont
  {D.}~\bibnamefont {Adroja}}, \bibinfo {author} {\bibfnamefont
  {T.}~\bibnamefont {Perring}}, \bibinfo {author} {\bibfnamefont
  {W.}~\bibnamefont {Liu}}, \bibinfo {author} {\bibfnamefont {F.}~\bibnamefont
  {Jin}}, \bibinfo {author} {\bibfnamefont {J.}~\bibnamefont {Ji}}, \bibinfo
  {author} {\bibfnamefont {Y.}~\bibnamefont {Wang}},  \emph {et~al.},\
  }\bibfield  {title} {\enquote {\bibinfo {title} {{Crystalline Electric-Field
  Excitations in Quantum Spin Liquids Candidate NaYbSe$_2$}},}\ }\href@noop {}
  {\bibfield  {journal} {\bibinfo  {journal} {arXiv preprint arXiv:2002.04772}\
  } (\bibinfo {year} {2020}{\natexlab{a}})}\BibitemShut {NoStop}%
\bibitem [{\citenamefont {Bordelon}\ \emph {et~al.}(2019)\citenamefont
  {Bordelon}, \citenamefont {Kenney}, \citenamefont {Hogan}, \citenamefont
  {Posthuma}, \citenamefont {Kavand}, \citenamefont {Lyu}, \citenamefont
  {Sherwin}, \citenamefont {Brown}, \citenamefont {Graf}, \citenamefont
  {Balents} \emph {et~al.}}]{bordelon2019field}%
  \BibitemOpen
  \bibfield  {author} {\bibinfo {author} {\bibfnamefont {M.}~\bibnamefont
  {Bordelon}}, \bibinfo {author} {\bibfnamefont {E.}~\bibnamefont {Kenney}},
  \bibinfo {author} {\bibfnamefont {T.}~\bibnamefont {Hogan}}, \bibinfo
  {author} {\bibfnamefont {L.}~\bibnamefont {Posthuma}}, \bibinfo {author}
  {\bibfnamefont {M.}~\bibnamefont {Kavand}}, \bibinfo {author} {\bibfnamefont
  {Y.}~\bibnamefont {Lyu}}, \bibinfo {author} {\bibfnamefont {M.}~\bibnamefont
  {Sherwin}}, \bibinfo {author} {\bibfnamefont {C.}~\bibnamefont {Brown}},
  \bibinfo {author} {\bibfnamefont {M.}~\bibnamefont {Graf}}, \bibinfo {author}
  {\bibfnamefont {L.}~\bibnamefont {Balents}},  \emph {et~al.},\ }\bibfield
  {title} {\enquote {\bibinfo {title} {{Field-tunable quantum disordered ground
  state in the triangular lattice antiferromagnet NaYbO$_2$}},}\ }\href
  {https://arxiv.org/abs/1901.09408} {\bibfield  {journal} {\bibinfo  {journal}
  {arXiv preprint arXiv:1901.09408}\ } (\bibinfo {year} {2019})}\BibitemShut
  {NoStop}%
\bibitem [{\citenamefont {Dai}\ \emph {et~al.}(2021)\citenamefont {Dai},
  \citenamefont {Zhang}, \citenamefont {Xie}, \citenamefont {Duan},
  \citenamefont {Gao}, \citenamefont {Zhu}, \citenamefont {Feng}, \citenamefont
  {Tao}, \citenamefont {Huang}, \citenamefont {Cao} \emph
  {et~al.}}]{dai2021spinon}%
  \BibitemOpen
  \bibfield  {author} {\bibinfo {author} {\bibfnamefont {P.-L.}\ \bibnamefont
  {Dai}}, \bibinfo {author} {\bibfnamefont {G.}~\bibnamefont {Zhang}}, \bibinfo
  {author} {\bibfnamefont {Y.}~\bibnamefont {Xie}}, \bibinfo {author}
  {\bibfnamefont {C.}~\bibnamefont {Duan}}, \bibinfo {author} {\bibfnamefont
  {Y.}~\bibnamefont {Gao}}, \bibinfo {author} {\bibfnamefont {Z.}~\bibnamefont
  {Zhu}}, \bibinfo {author} {\bibfnamefont {E.}~\bibnamefont {Feng}}, \bibinfo
  {author} {\bibfnamefont {Z.}~\bibnamefont {Tao}}, \bibinfo {author}
  {\bibfnamefont {C.-L.}\ \bibnamefont {Huang}}, \bibinfo {author}
  {\bibfnamefont {H.}~\bibnamefont {Cao}},  \emph {et~al.},\ }\bibfield
  {title} {\enquote {\bibinfo {title} {{Spinon Fermi Surface Spin Liquid in a
  Triangular Lattice Antiferromagnet NaYbSe$_2$}},}\ }\href@noop {} {\bibfield
  {journal} {\bibinfo  {journal} {Physical Review X}\ }\textbf {\bibinfo
  {volume} {11}},\ \bibinfo {pages} {021044} (\bibinfo {year}
  {2021})}\BibitemShut {NoStop}%
\bibitem [{\citenamefont {Xie}\ \emph {et~al.}(2021)\citenamefont {Xie},
  \citenamefont {Xing}, \citenamefont {Nikitin}, \citenamefont {Nishimoto},
  \citenamefont {Brando}, \citenamefont {Khanenko}, \citenamefont
  {Sichelschmidt}, \citenamefont {Sanjeewa}, \citenamefont {Sefat},\ and\
  \citenamefont {Podlesnyak}}]{xie2021field}%
  \BibitemOpen
  \bibfield  {author} {\bibinfo {author} {\bibfnamefont {T.}~\bibnamefont
  {Xie}}, \bibinfo {author} {\bibfnamefont {J.}~\bibnamefont {Xing}}, \bibinfo
  {author} {\bibfnamefont {S.}~\bibnamefont {Nikitin}}, \bibinfo {author}
  {\bibfnamefont {S.}~\bibnamefont {Nishimoto}}, \bibinfo {author}
  {\bibfnamefont {M.}~\bibnamefont {Brando}}, \bibinfo {author} {\bibfnamefont
  {P.}~\bibnamefont {Khanenko}}, \bibinfo {author} {\bibfnamefont
  {J.}~\bibnamefont {Sichelschmidt}}, \bibinfo {author} {\bibfnamefont
  {L.}~\bibnamefont {Sanjeewa}}, \bibinfo {author} {\bibfnamefont {A.~S.}\
  \bibnamefont {Sefat}}, \ and\ \bibinfo {author} {\bibfnamefont
  {A.}~\bibnamefont {Podlesnyak}},\ }\bibfield  {title} {\enquote {\bibinfo
  {title} {{Field-Induced Spin Excitations in the Spin-1/2 Triangular-Lattice
  Antiferromagnet CsYbSe$_2$}},}\ }\href@noop {} {\bibfield  {journal}
  {\bibinfo  {journal} {arXiv preprint arXiv:2106.12451}\ } (\bibinfo {year}
  {2021})}\BibitemShut {NoStop}%
\bibitem [{\citenamefont {Maksimov}\ \emph {et~al.}(2019)\citenamefont
  {Maksimov}, \citenamefont {Zhu}, \citenamefont {White},\ and\ \citenamefont
  {Chernyshev}}]{maksimov2019anisotropic}%
  \BibitemOpen
  \bibfield  {author} {\bibinfo {author} {\bibfnamefont {P.}~\bibnamefont
  {Maksimov}}, \bibinfo {author} {\bibfnamefont {Z.}~\bibnamefont {Zhu}},
  \bibinfo {author} {\bibfnamefont {S.~R.}\ \bibnamefont {White}}, \ and\
  \bibinfo {author} {\bibfnamefont {A.}~\bibnamefont {Chernyshev}},\ }\bibfield
   {title} {\enquote {\bibinfo {title} {{Anisotropic-exchange magnets on a
  triangular lattice: spin waves, accidental degeneracies, and dual spin
  liquids}},}\ }\href@noop {} {\bibfield  {journal} {\bibinfo  {journal}
  {Physical Review X}\ }\textbf {\bibinfo {volume} {9}},\ \bibinfo {pages}
  {021017} (\bibinfo {year} {2019})}\BibitemShut {NoStop}%
\bibitem [{\citenamefont {Yamamoto}, \citenamefont {Marmorini},\ and\
  \citenamefont {Danshita}(2014)}]{yamamoto2014quantum}%
  \BibitemOpen
  \bibfield  {author} {\bibinfo {author} {\bibfnamefont {D.}~\bibnamefont
  {Yamamoto}}, \bibinfo {author} {\bibfnamefont {G.}~\bibnamefont {Marmorini}},
  \ and\ \bibinfo {author} {\bibfnamefont {I.}~\bibnamefont {Danshita}},\
  }\bibfield  {title} {\enquote {\bibinfo {title} {{Quantum Phase Diagram of
  the Triangular-Lattice $XXZ$ Model in a Magnetic Field}},}\ }\href {\doibase
  10.1103/PhysRevLett.112.127203} {\bibfield  {journal} {\bibinfo  {journal}
  {Phys. Rev. Lett.}\ }\textbf {\bibinfo {volume} {112}},\ \bibinfo {pages}
  {127203} (\bibinfo {year} {2014})}\BibitemShut {NoStop}%
\bibitem [{\citenamefont {Jia}\ \emph {et~al.}(2020)\citenamefont {Jia},
  \citenamefont {Gong}, \citenamefont {Liu}, \citenamefont {Zhao},
  \citenamefont {Dong}, \citenamefont {Dai}, \citenamefont {Li}, \citenamefont
  {Lei}, \citenamefont {Yu}, \citenamefont {Zhang} \emph
  {et~al.}}]{jia2020mott}%
  \BibitemOpen
  \bibfield  {author} {\bibinfo {author} {\bibfnamefont {Y.-T.}\ \bibnamefont
  {Jia}}, \bibinfo {author} {\bibfnamefont {C.-S.}\ \bibnamefont {Gong}},
  \bibinfo {author} {\bibfnamefont {Y.-X.}\ \bibnamefont {Liu}}, \bibinfo
  {author} {\bibfnamefont {J.-F.}\ \bibnamefont {Zhao}}, \bibinfo {author}
  {\bibfnamefont {C.}~\bibnamefont {Dong}}, \bibinfo {author} {\bibfnamefont
  {G.-Y.}\ \bibnamefont {Dai}}, \bibinfo {author} {\bibfnamefont {X.-D.}\
  \bibnamefont {Li}}, \bibinfo {author} {\bibfnamefont {H.-C.}\ \bibnamefont
  {Lei}}, \bibinfo {author} {\bibfnamefont {R.-Z.}\ \bibnamefont {Yu}},
  \bibinfo {author} {\bibfnamefont {G.-M.}\ \bibnamefont {Zhang}},  \emph
  {et~al.},\ }\bibfield  {title} {\enquote {\bibinfo {title} {{Mott transition
  and superconductivity in quantum spin liquid candidate NaYbSe$_2$}},}\
  }\href@noop {} {\bibfield  {journal} {\bibinfo  {journal} {Chinese Physics
  Letters}\ }\textbf {\bibinfo {volume} {37}},\ \bibinfo {pages} {097404}
  (\bibinfo {year} {2020})}\BibitemShut {NoStop}%
\bibitem [{\citenamefont {Zhang}\ \emph
  {et~al.}(2020{\natexlab{b}})\citenamefont {Zhang}, \citenamefont {Yin},
  \citenamefont {Ma}, \citenamefont {Liu}, \citenamefont {Li}, \citenamefont
  {Jin}, \citenamefont {Ji}, \citenamefont {Wang}, \citenamefont {Wang},
  \citenamefont {Yu} \emph {et~al.}}]{zhang2020pressure}%
  \BibitemOpen
  \bibfield  {author} {\bibinfo {author} {\bibfnamefont {Z.}~\bibnamefont
  {Zhang}}, \bibinfo {author} {\bibfnamefont {Y.}~\bibnamefont {Yin}}, \bibinfo
  {author} {\bibfnamefont {X.}~\bibnamefont {Ma}}, \bibinfo {author}
  {\bibfnamefont {W.}~\bibnamefont {Liu}}, \bibinfo {author} {\bibfnamefont
  {J.}~\bibnamefont {Li}}, \bibinfo {author} {\bibfnamefont {F.}~\bibnamefont
  {Jin}}, \bibinfo {author} {\bibfnamefont {J.}~\bibnamefont {Ji}}, \bibinfo
  {author} {\bibfnamefont {Y.}~\bibnamefont {Wang}}, \bibinfo {author}
  {\bibfnamefont {X.}~\bibnamefont {Wang}}, \bibinfo {author} {\bibfnamefont
  {X.}~\bibnamefont {Yu}},  \emph {et~al.},\ }\bibfield  {title} {\enquote
  {\bibinfo {title} {{Pressure induced metallization and possible
  unconventional superconductivity in spin liquid NaYbSe$_2$}},}\ }\href@noop
  {} {\bibfield  {journal} {\bibinfo  {journal} {arXiv preprint
  arXiv:2003.11479}\ } (\bibinfo {year} {2020}{\natexlab{b}})}\BibitemShut
  {NoStop}%
\bibitem [{\citenamefont {Gray}, \citenamefont {Martin},\ and\ \citenamefont
  {Dorhout}(2003)}]{gray2003crystal}%
  \BibitemOpen
  \bibfield  {author} {\bibinfo {author} {\bibfnamefont {A.}~\bibnamefont
  {Gray}}, \bibinfo {author} {\bibfnamefont {B.}~\bibnamefont {Martin}}, \ and\
  \bibinfo {author} {\bibfnamefont {P.}~\bibnamefont {Dorhout}},\ }\bibfield
  {title} {\enquote {\bibinfo {title} {{Crystal structure of potassium
  ytterbium (III) selenide, KYbSe$_2$}},}\ }\href@noop {} {\bibfield  {journal}
  {\bibinfo  {journal} {Zeitschrift f{\"u}r Kristallographie-New Crystal
  Structures}\ }\textbf {\bibinfo {volume} {218}},\ \bibinfo {pages} {20--20}
  (\bibinfo {year} {2003})}\BibitemShut {NoStop}%
\bibitem [{\citenamefont {Sheldrick}(2015)}]{sheldrick2015crystal}%
  \BibitemOpen
  \bibfield  {author} {\bibinfo {author} {\bibfnamefont {G.~M.}\ \bibnamefont
  {Sheldrick}},\ }\bibfield  {title} {\enquote {\bibinfo {title} {{Crystal
  structure refinement with SHELXL}},}\ }\href@noop {} {\bibfield  {journal}
  {\bibinfo  {journal} {Acta Crystallographica Section C: Structural
  Chemistry}\ }\textbf {\bibinfo {volume} {71}},\ \bibinfo {pages} {3--8}
  (\bibinfo {year} {2015})}\BibitemShut {NoStop}%
\end{thebibliography}%

\end{document}